\documentclass[english,preprint,aps,prd,showpacs,superscriptaddress,nofootinbib,tightenlines]{revtex4}
\usepackage[T1]{fontenc}
\usepackage[latin9]{inputenc}
\usepackage{float}
\usepackage{amsmath}
\usepackage{graphicx}
\usepackage{esint}
\usepackage{ulem}
\usepackage{amsfonts}
\usepackage{multirow}
\usepackage{mathrsfs}
\usepackage{graphicx}
\usepackage{amsmath}
\usepackage{amssymb}
\usepackage{bm}
\usepackage{bbm}
\usepackage{bm}
\usepackage{xcolor}
\usepackage{babel}
\usepackage{float}

\newcommand{\bseq}{\begin{subequations}}

\newcommand{\eseq}{\end{subequations}}
\newcommand{\emphs}[1]{\textcolor{black}{#1}}
\newcommand{\emphz}[1]{\textcolor{black}{#1}}
\usepackage{scalerel,stackengine}
\usepackage{babel}

\stackMath
\renewcommand\widehat[1]{%
\savestack{\tmpbox}{\stretchto{%
  \scaleto{%
    \scalerel*[\widthof{\ensuremath{#1}}]{\kern-.6pt\bigwedge\kern-.6pt}%
    {\rule[-\textheight/2]{1ex}{\textheight}}
  }{\textheight}%
}{0.5ex}}%
\stackon[1pt]{#1}{\tmpbox}%
}
\renewcommand\widetilde[1]{%
\savestack{\tmpbox}{\stretchto{%
  \scaleto{%
    \scalerel*[\widthof{\ensuremath{#1}}]{\kern-.6pt\boldsymbol{\sim}\kern-.6pt}%
    {\rule[-\textheight/2]{1ex}{\textheight}}
  }{\textheight}%
}{0.5ex}}%
\stackon[1pt]{#1}{\tmpbox}%
}

\begin{document}

\title{Quasi-Parton Distribution Function in Lattice Perturbation Theory}

\author{Xiaonu Xiong\footnote{x.xiong@fz-juelich.de}}
\affiliation{Institute for Advanced Simulation, Institut f\"ur Kernphysik and J\"ulich Center for Hadronphysics,
Forschungszentrum J\"ulich, D-52428 J\"ulich, Germany\vspace{0.2cm}}

\author{Thomas Luu\footnote{t.luu@fz-juelich.de}}
\affiliation{Institute for Advanced Simulation, Institut f\"ur Kernphysik and J\"ulich Center for Hadronphysics,
Forschungszentrum J\"ulich, D-52428 J\"ulich, Germany\vspace{0.2cm}}
\affiliation{JARA -High Performance Computing, Forschungszentrum J\"ulich, D-52428 J\"ulich, Germany\vspace{0.2cm}}

\author{Ulf-G. Mei{\ss}ner\footnote{meissner@hiskp.uni-bonn.de}}
\affiliation{Helmholtz-Institut f\"ur Strahlen- und Kernphysik and Bethe Center for Theoretical Physics, Universit\"at Bonn, D-53115 Bonn, Germany}
\affiliation{Institute for Advanced Simulation, Institut f\"ur Kernphysik and J\"ulich Center for Hadronphysics,
Forschungszentrum J\"ulich, D-52428 J\"ulich, Germany\vspace{0.2cm}}
\affiliation{JARA -High Performance Computing, Forschungszentrum J\"ulich, D-52428 J\"ulich, Germany\vspace{0.2cm}}

\date{\today}

\begin{abstract}
Large momentum effective field theory provides a new direction
for lattice QCD calculations of hadronic structure functions, such as
parton distribution functions (PDFs), meson distribution amplitudes, and so on, directly
with $x$-dependence. In the framework of Lattice Perturbation Theory (LPT),
we compute the one-loop quark-in-quark quasi-PDF with the na\"ive fermion action ($\tilde{q}^{\mathrm{nv}}$) and quasi-PDF with Wilson-Clover action ($\tilde{q}^{\mathrm{WC}}$) and show that $\tilde{q}^{\mathrm{nv}}$ reduces to the
continuum quasi-PDF in the continuum limit.  We point out, however, that the continuum limit
and massless quark limit do not commute. We find that the condition to recover the same collinear divergence that the quasi-PDF has in continuum QCD is $aP_3^2\approx m$ and $m\ll P_3$, while the condition to fully recover the continuum quasi-PDF is $aP_{3}^{2}\ll m\ll P_{3}$, where $P_3$ is the momentum in the direction of the quark's motion (longitudinal direction). These two conditions are based on perturbation calculations and should not be applied to non-perturbative calculations because the non-perturbative effects cure the collinear divergence. The correction
to the quasi-PDF using the na\"ive fermion action is due to the Wilson term and can be viewed as an $\mathcal{O}\left(a^{1}\right)$ correction.
For nonzero $r$, the $\mathcal{O}\left(a^{1}\right)$ corrections are subsequently mixed with the quasi-PDF using the na\"ive fermion action.
\end{abstract}



\maketitle

\section{introduction}

The parton distribution functions (PDFs) provide essential information for understanding various aspects of  the internal structure of hadrons, such
as the nucleon spin structure~\cite{Ji:1996ek,Ji:2012sj,Yang:2016plb} and the
flavor structure of the proton~\cite{Peng:2014uea,Chang:2014jba}. Thanks to the concept of factorization, which separates the perturbatively calculable short-range processes and the complex long-range behavior, the latter one is absorbed into PDFs~\cite{Collins:1989gx}. PDFs also can be widely used as important inputs for high-energy experiments involving hadrons that, for example, probe new physics at hadron colliders~\cite{Martin:2009iq,Lai:2010vv}. The experimental determination and application of PDFs are based on their universality, which allows one to extract the PDFs from various types of high-energy scattering processes  measured in different types of experiments. The factorization and universality of the PDFs play a central role in QCD predictions. Therefore, the determination of PDFs has been a long-standing key task in QCD. Although PDFs can be measured experimentally, there are still some regions that experiments can not  cover or present large uncertainties (e.g. PDFs at small and large Bjorken $x$ and gluon PDFs)~\cite{Dulat:2015mca,Gauld:2016kpd}. Theoretical studies can help  provide more precise and complete PDFs. Due to the  non-perturbative nature of hadrons, the theoretical study of their internal structure requires non-perturbative methods. Previous studies of hadronic PDFs used QCD models and Ads/CFT QCD. However, the model dependence in those studies has limited their predictive power. Lattice QCD provides so far the only reliable first-principle non-perturbative QCD method, but calculating PDFs through lattice QCD has its intrinsic difficulties.

The light-cone quark PDFs are defined via~\cite{Collins:1981uw}
\begin{equation}
q_{\Gamma}\left(x\right)=\int\frac{d\xi^{-}}{4\pi}\,e^{-ixP^{+}\xi^{-}}\left\langle P\left|\bar{\psi}\left(\xi^{-}\right)\Gamma\mathcal{P}\left\{ \exp\left[-ig\int_{0}^{\xi^{-}}d\eta^{-}A^{+}(\eta^{-})\right]\right\} \psi\left(0\right)\right|P\right\rangle~,
\end{equation}
where $\Gamma=\gamma^{+},\gamma^{+}\gamma_{5},\gamma^{+}\gamma^{\perp}$ and the light-cone components of a vector $v^\mu$ are {$v^\pm=\left(v^0\pm v^z\right)/\sqrt{2}$, $v^\perp=v^{1,2}$.} The different $\gamma$-matrices
correspond to the unpolarized PDF, helicity distribution function
and transversity distribution function, respectively. The path-ordered exponential
is the gauge link, which ensures the gauge invariance of the non-local quark correlator. The
light-cone PDFs are based on light-cone correlation functions while
 lattice QCD correlators are evaluated in Euclidean space-time.    The field separation
$\xi^{-}$ becomes complex in Euclidean space-time. Consequently, lattice
QCD can not calculate  the explicit $x$-dependence of any PDF directly.
Instead, lattice QCD calculations focus on the PDF's Mellin
moments which reduce to local operator matrix elements
\begin{equation}
q^{n}=\int_{0}^{1}dx\,x^{n-1}q\left(x\right)\backsim\left\langle P\left|\bar{\psi}\left(0\right)\gamma^{\{+}\overleftrightarrow{D}^{+}\cdots\overleftrightarrow{D}^{+\}}\psi\left(0\right)\right|P\right\rangle-\rm{trace}~.\label{eq:x_Moments}
\end{equation}
These can be calculated directly on the lattice. The PDFs are then reconstructed
from these Mellin moments. However, due to operator mixing and discretization errors, the high moments are very difficult
to calculate.

The proposed large momentum effective field theory provides the ability
to calculate PDFs with their $x$-dependence directly on the lattice. In
the large momentum effective field theory, the light-cone PDFs are
accessible from particular pure spatial correlation functions: the
so-called quasi-PDFs, which are defined as~\cite{Ji:2013dva,Ji:2014gla}
\begin{equation}\label{eq:QSPDF_Dfntn}
\tilde{q}_{\tilde{\Gamma}}(x,\mu^{2},P^{z})=\int\frac{dz}{4\pi}e^{ixP^{z}z}\left\langle P\left|\overline{\psi}(z)\tilde{\Gamma}\mathcal{P}\left\{ \exp\left[-ig\int_{0}^{z}dz'A^{z}(z')\right]\right\} \psi(0)\right|P\right\rangle ,
\end{equation}
where again $\tilde{\Gamma}=\gamma^{z},\gamma^{z}\gamma_{5},\gamma^{z}\gamma^{\perp}$. The gauge link lies in the $z$-direction. The fact that the field separation $z$ is purely spatial and no longer complex enables such a quantity to be calculated
directly on the lattice. The quasi-PDFs and light-cone PDFs are then related
through a matching condition~\cite{Xiong:2013bka}
\begin{equation}
\tilde{q}\left(x\right)=\int_{-1}^{1}\frac{dy}{\left|y\right|}\,Z\left(\frac{x}{y}\right)q\left(y\right)+\mathcal{O}\left(\frac{M_{N}^{n}}{\left(P^{z}\right)^{n}}\right)+\mathcal{O}\left(\frac{\varLambda_{QCD}^{n}}{\left(P^{z}\right)^{n}}\right)~,
\label{eq:matchhing_condition}
\end{equation}
where $Z$ is the matching factor and the last two terms refer to the target mass and higher twist corrections, respectively. The proof of the above matching condition can be found in Refs.~\cite{Ma:2014jla,Ma:2014jga}. The target
mass corrections of quark PDFs has been studied in Ref.~\cite{Chen:2016utp}.
The quasi-PDFs are assumed to have the same infrared behavior as the
light-cone PDFs, and under this assumption, the matching factor $Z$ is totally controlled
by the ultraviolet (UV) behavior of quasi-PDF and light-cone PDF.
As a consequence, the matching factor can be calculated by perturbation
theory. The matching factor $Z$ is calculated up to $\mathcal{O}\left(\alpha_{s}\right)$
in continuum QCD in Refs.~\cite{Xiong:2013bka,Ma:2014jla} and the
renormalization of quasi-PDFs up to $\mathcal{O}\left(\alpha_{s}^{2}\right)$
is studied in Ref.~\cite{Ji:2015jwa}. The matching scheme in position
space is studied in Ref.~\cite{Ishikawa:2016znu}. Some possible improved
definitions of quasi-PDFs are suggested in Refs.~\cite{Chen:2016fxx,Li:2016amo}.
The matching for the PDF's generalizations, Generalized Parton Distributions
(GPDs), have been studied in Refs.~\cite{Ji:2015qla,Xiong:2015nua}. Lattice simulations of quasi-unpolarized
parton distributions, quark helicity distributions and transversity
distributions have been performed in Refs.~\cite{Lin:2014zya,Alexandrou:2015rja,Chen:2016utp}. LPT calculations of the quasi-PDF
using Wilson fermions up to $\mathcal{O}\left(\alpha_{s}^{1},a^0\right)$
can be found in Ref.~\cite{Carlson:2017gpk}. In this work they discuss the discrepancy of the IR behavior between Euclidean and Minkowski quasi-PDFs. Ref.~\cite{Briceno:2017cpo} provides a solution to this discrepancy. \emphz{Recently, the perturbative and nonperturbative renormalization of quasi-PDFs on the lattice have been studied in Refs.~\cite{Constantinou:2017sej,Alexandrou:2017huk} using the RI scheme and lattice simulations of quasi-PDFs with renormalization in the RI/MOM scheme is presented in Ref.~\cite{Chen:2017mzz}.}

In this work we focus
on calculating the one-loop quark-in-quark quasi-PDF (unpolarized) from lattice perturbation
theory. We use both the na\"ive fermion and Wilson-Clover fermion actions as our discretization. We find that, due to lattice artifacts, the collinear divergence is absent in the LPT-calculated quasi-PDF when the lattice spacing is not too small, even after extending the LPT-calculated quasi-PDFs to Minkowski space. Furthermore, since the massless quark limit and the continuum limit do not commute \emphz{in the LPT-calculated quasi-PDFs, the exchange of these two limits leads to different IR (collinear) behavior of the quasi-PDF in LPT at finite lattice spacing}. The correct condition that reproduces the same collinear divergence in the continuum limit quasi-PDF is $aP_3^2\approx m$ and $m \ll P_3$. The limit for fully reproducing the quasi-PDF in continuum QCD is $aP_3^2\ll m \ll P_3$. These two conditions should be constrained to perturbative calculations, because the non-perturbative effects in non-pertrubative lattice calculations remove the collinear divergence. We also observe that the $\mathcal{O}\left(a^1\right)$ corrections turn out to be mixed with the $\mathcal{O}\left(a^0\right)$ quasi-PDF. The presence of this mixing limits the lattice perturbation calculation to a ballpark estimation of the matching factor between lattice and continuum.  As such, a non-perturbative matching is more appropriate and will be studied in future work.

The rest of the paper
is organized as follows: in Sec.~II we introduce the Wilson-Clover
fermion action and na\"ive fermion action, as well as their corresponding Feynman rules. In Sec.~III we
describe the calculation method and present the analytical result
of the quasi-PDF in LPT with the na\"ive fermion action, and show that the quasi-PDF is an integral over transverse momentum
$\boldsymbol{k}_{\perp}$. We also compare the $a\rightarrow0$ limit
of our result and the quasi-PDF directly calculated in continuum QCD
and discuss the collinear divergence term in the two quasi-PDFs. {We
will also shortly present the calculation of quasi-PDF in LPT with Wilson-Clover fermon action}. In Sec.~IV
we present our numerical results for the  quasi-PDF in LPT
and its comparison with the continuum quasi-PDF. We conclude in Sec.~V.

\section{na\"ive and wilson-clover fermion action and feynman rules}

The Wilson-Clover action is given by
\begin{align}
S= & -\frac{1}{2}\sum_{x,\mu}\left[\bar{\psi}\left(x\right)\left(r-\gamma_{\mu}\right)U_{\mu}\left(x\right)\psi\left(x+a\hat{\mu}\right)+\bar{\psi}\left(x+\mu\right)\left(r+\gamma_{\mu}\right)U_{\mu}^{\dagger}\left(x-a\hat{\mu}\right)\psi\left(x\right)\right]\nonumber \\
 & +\sum_{x}\left[\left(4r+m\right)\bar{\psi}\left(x\right)\psi\left(x\right)-\sum_{\mu,\nu}c_{SW}g_{s}\frac{a}{4}\bar{\psi}\left(x\right)\sigma_{\mu\nu}\hat{F}_{\mu\nu}\left(x\right)\psi\left(x\right)\right],
\end{align}
where the gauge field strength tensor is defined as
\begin{align}
\hat{F}_{\mu\nu}\equiv\frac{1}{8}\left(Q_{\mu\nu}-Q_{\nu\mu}\right),
\end{align}
and $Q_{\mu\nu}$ denotes the sum of plaquette loops. \emphz{We use the plaquette gauge action for gluon.} The Wilson-Clover
fermion action violates chiral symmetry explicitly through the terms
proportional to $r$. We choose Wilson-Clover fermions because
we focus only on the unpolarized parton distribution functions and such terms do not depend crucially on chiral symmetry (and breaking thereof).
Further, Wilson-Clover fermions provide a reasonable computational cost for
most practical lattice calculations.

The Wilson-Clover fermion's on-shell
condition can be solved from the lattice discretized Dirac equation
\begin{equation}
\left[i\sum_{\mu}\gamma_{\mu}\widehat{2P}_{\mu}+r\sum_{\mu}\left(\frac{2}{a}-\widetilde{2P}_{\mu}\right)+2m\right]U\left(P\right)=0,
\end{equation}
where the lattice momenta are
\begin{equation}
\widehat{P}_{\mu}=\frac{2}{a}\sin\frac{aP_{\mu}}{2},\;\;\widetilde{P}_{\mu}=\frac{2}{a}\cos\frac{aP_{\mu}}{2}~.
\end{equation}
The Dirac equation gives the dispersion relation for the Wilson-Clover
fermions
\begin{align}\label{eq:OnShell}
P_{4}= & \frac{1}{a}\sinh^{-1}\left(\frac{1}{\sqrt{2}}\left\{ \frac{1}{\left(1-r^{2}\right)^{2}}\left[2\left(r^{2}+1\right)(am+2r)\left(a^{2}r\widehat{P}_{3}^{2}+am\right)\right.\right.\right.\nonumber \\
\notag & -\frac{1}{2}a^{2}\left(r^{4}+2r^{2}-1\right)\widehat{2P}_{3}^{2}+4r^{2}-\left(a^{2}r^{2}\widehat{P}_{3}^{2}+2ram+2r^{2}\right)\\
& \left.\left.\left.\times\sqrt{a^{4}\left(2r^{2}-1\right)\widehat{P}_{3}^{4}+4a^{2}\widehat{P}_{3}^{2}(amr+1)+4am(am+2r)+4}\,\right]\right\} ^{\frac{1}{2}}\right)~,
\end{align}
in which we have set the quark moving along the $z-$direction $P_{\mu}=\left(0,0,P_{3},P_{4}\right)$ {and we keep this momentum setup for the parent quark throughout the paper.}
In the continuum limit, this reduces to the conventional on-shell condition
\begin{equation}
\lim_{a\rightarrow0}P_{4}=\sqrt{P_{3}^{2}+m^{2}}.
\end{equation}

The Feynman rules for Wilson-Clover fermion action can be found in
Ref.~\cite{Horsley:2008ap}. The quark propagator is given by
\begin{align}
S_{F}\left(k\right)= & 2\left[\frac{-i\displaystyle{\sum_{\mu}}\gamma_{\mu}\widehat{2k}_{\mu}+r\sum_{\mu}\left(\frac{2}{a}-\widetilde{2k}_{\mu}\right)+2m}{\widehat{2k}^{2}+\left(r\displaystyle\sum_{\mu}\left(\frac{2}{a}-\widetilde{2k}_{\mu}\right)+2m\right)^{2}}\right],
\label{eq:S_F}
\end{align}
where $\widehat{k}^{2}=\sum_{\mu}\widehat{k}_{\mu}^{2}$.

The complete form of the gluon propagator in the covariant gauge is quite complicated but can be found in Refs.~\cite{Weisz:1982zw,Weisz:1983bn}. In this work we only need the gluon propagator up to order $\mathcal{O}\left(a^{1}\right)$
\begin{align}
D_{g,\mu\nu}\left(k\right)= & \frac{1}{\widehat{k}^{2}}\left[\delta_{\mu\nu}-\left(1-\xi\right)\frac{a^{2}}{4}\widehat{k}_{\mu}\widehat{k}_{\nu}\right].
\end{align}
We choose Feynman gauge $\xi=1$ in this work. The quark-gluon-quark
interaction vertex is given by{~\cite{Horsley:2008ap}}
\begin{align}
V_{\alpha}^{a}\!\left(p_{2},p_{1},k\right)= \!-\!ig_{s}T^{a}\frac{a}{2}\!\left(\widetilde{p_{2}\!+\!p_{1}}\right)_{\alpha}\!\!\gamma_{\alpha}\!
\!-\!g_{s}T^{a}r\frac{a}{2}\!\left(\widehat{p_{2}\!+\!p_{1}}\right)_{\alpha}\!\!-\!ig_{s}T^{a}rc_{\mathrm{SW}}
\frac{a^{2}}{8}\tilde{k}_{\alpha}\!\sum_{\mu}\!\sigma_{\alpha\mu}\widehat{2k}_{\mu}~,
\end{align}
where $p_{2}=p_{1}+k$, the fermion momenta directions are
assigned parallel to the direction of fermion line and $k$ is the
momentum of the gluon.

\emphs{The na\"ive fermion action and Feynman rules are given by setting $r\rightarrow0$ in the Wilson-Clover action and Feynman rules.}

\section{one-loop corrections for quasi-PDF}

We now present the procedure for calculating within LPT
the quasi-PDF in detail and the resulting quasi-PDF will be provided
as an integrand of $\boldsymbol{k}_{\perp}$ in analytical form to order $\mathcal{O}\left(a^{0}\right)$, and as an integrand of $\boldsymbol{k}_{\perp},\,k_{4}$ to
order $\mathcal{O}\left(a^{1}\right)$. We will discuss the collinear
behavior of the quasi-PDF determined from LPT. As this work only
concentrates on the unpolarized quark-in-quark PDF, quark-gluon mixing
is not considered. The relevant Feynman diagrams at order $\mathcal{O}\left(\alpha_{s}a^{0}\right)+\mathcal{O}\left(\alpha_{s}a^{1}\right)$
are shown in Fig.~\ref{fig:One-loop-real-correction}. We have
omitted those virtual correction diagrams (shown in Fig.~\ref{fig:One-loop-virtual-correction}) in our calculation, because their contributions
to the quasi-PDF are proportional to $\delta\left(x-1\right)$ and hence only
contribute at $x=1$.
\begin{center}
\begin{figure}[H]
\centering{}\includegraphics[width=\textwidth]{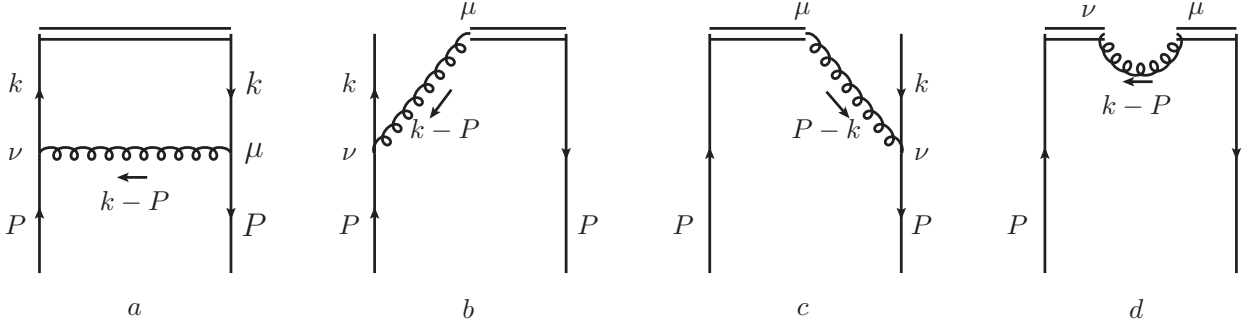}\caption{One-loop correction diagrams, the  double line represents the gauge link in the quasi-PDF definition \eqref{eq:QSPDF_Dfntn}.  \label{fig:One-loop-real-correction}}
\end{figure}
\par\end{center}
\begin{center}
\begin{figure}[H]
\centering{}\includegraphics[width=0.975\textwidth]{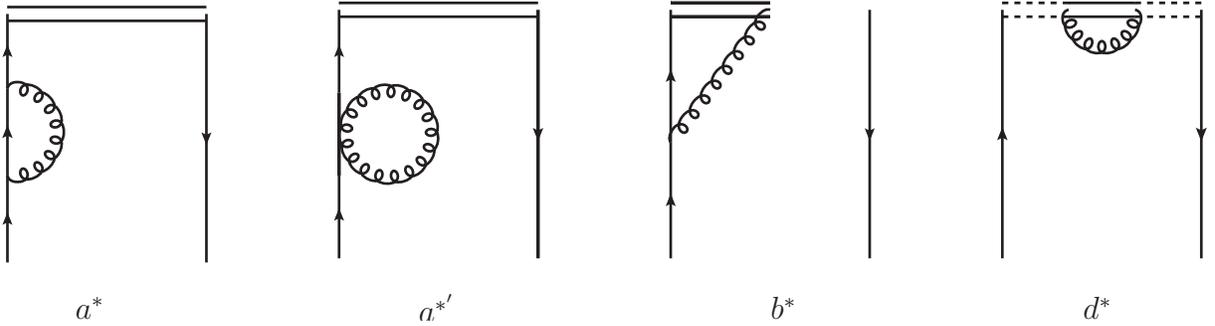}\caption{One-loop virtual correction diagrams of the quasi-PDF. They contribute to the quasi-PDF as $Z\delta(x-1)$, where $Z$ is a $x$-independent constant. The quark self-energy diagram $a^*$ (sunset) and $a^{*'}$ (tadpole) will contribute to the additive mass renormalization for Wilson fermions.}
\label{fig:One-loop-virtual-correction}
\end{figure}
\par\end{center}

The one-loop Feynman diagrams lead to four terms,\footnote{The one-loop corrections will 
induce a scalar fermion bilinear structure $\bar{U}\left(P\right)U\left(P\right)$, 
which can be related to the tree-level quasi-PDF fermion bilinear structure by 
$\displaystyle{\bar{U}\left(P\right)U\left(P\right)=\varDelta = 
\frac{\varDelta}{\mathcal{P}_3}\bar{U}\left(P\right)\gamma_3 U\left(P\right)}$, in which 
$\mathcal{P}_\mu$ and $\varDelta$ is defined in Eq.~\eqref{eq:calP_varDlt} (the continuum 
limit correspondence $\displaystyle{\bar{u}\left(P\right)u\left(P\right)=2m = 
\frac{m}{P^3}\bar{u}\left(P\right)\gamma^3 u\left(P\right)}$  has been used in 
Ref.~\cite{Xiong:2013bka}). We see that this scalar bilinear term's contribution is 
at higher twist. The mixing is included by taking the trace of numerator and denominator 
on the right side of \eqref{eq:f0_abcd}.}
\bseq\label{eq:f0_abcd}
\begin{align}
\notag \tilde{q}_{a}\left(x\right)= & \int_{-\frac{\pi}{a}}^{\frac{\pi}{a}}\frac{d^{4}k}{\left(2\pi\right)^{4}}
\frac{\displaystyle\sum_{\mu,\nu}\,\bar{U}\left(P\right)V_{\mu}\left(P,k,P\!-\!k\right)S_{F}\left(k\right)
\gamma_{3}S_{F}\left(k\right)V_{\nu}\left(k,P,k\!-\!P\right)U\left(P\right)}{\bar{U}\left(P\right)
\gamma_{3}U\left(P\right)}\\
&\times D_{g,\mu\nu}\left(P\!-\!k\right)\delta\left(x\!-\!\frac{k^{3}}{P^{3}}\right),
\end{align}
\begin{align}
\notag \tilde{q}_{b}\left(x\right)= & \int_{-\frac{\pi}{a}}^{\frac{\pi}{a}}\frac{d^{4}k}{\left(2\pi\right)^{4}}\,
\frac{\displaystyle\sum_{\mu\nu}\bar{U}\left(P\right)O_{1,\mu}\left(P,k,P-k\right)S_{F}\left(k\right)
V_{\nu}\left(k,P,k-P\right)U\left(P\right)}{\bar{U}\left(P\right)\gamma_{3}U\left(P\right)}\\
&\times D_{g,\mu\nu}\left(P-k\right)\delta\left(x-\frac{k^{3}}{P^{3}}\right),
\end{align}
\begin{align}
\notag \tilde{q}_{c}\left(x\right)= & \int_{-\frac{\pi}{a}}^{\frac{\pi}{a}}\frac{d^{4}k}{\left(2\pi\right)^{4}}\,
\frac{\displaystyle\sum_{\mu\nu}\bar{U}\left(P\right)V_{\mu}\left(P,k,P-k\right)S_{F}\left(k\right)
O_{1,\nu}\left(k,P,k-P\right)U\left(P\right)}{\bar{U}\left(P\right)\gamma_{3}U\left(P\right)}\\
&\times D_{g,\mu\nu}\left(P-k\right)\delta\left(x-\frac{k^{3}}{P^{3}}\right),\\
\tilde{q}_{d}\left(x\right)= & \int_{-\frac{\pi}{a}}^{\frac{\pi}{a}}\frac{d^{4}k}{\left(2\pi\right)^{4}}\,
\frac{\displaystyle\sum_{\mu\nu}\bar{U}\left(P\right)O_{2,\mu\nu}\left(P,P,k-P\right)U\left(P\right)}{\bar{U}\left(P\right)
\gamma_{3}U\left(P\right)} D_{g,\mu\nu}\left(P-k\right)\delta\left(x-\frac{k^{3}}{P^{3}}\right).\label{eq:fd_intgrnd}
\end{align}
\eseq
The momentum space gluon and gauge-link interaction terms can
be obtained by Fourier transform of their corresponding coordinate expressions~\cite{Ishikawa:2016znu}, leading to
\bseq
\begin{align}
O_{1,\mu}^{A}\left(q\right)= & \frac{ig_{s}aT^{A}\gamma_{3}\delta_{\mu3}}{i\hat{q}_{3}},
\end{align}
\begin{align}
O_{2,\mu\nu}^{AB}\left(p,p,q\right)= & -g_{s}^{2}a^{2}\left\{ T^{A},T^{B}\right\} \gamma_{3}\delta_{\mu3}\delta_{\nu3}\frac{1}{\hat{q}_{3}^{2}}.
\end{align}
\eseq We note that there exists an extra $\mathcal{O}\left(a^{1}\right)$ terms in  coordinate space,
\begin{equation}
g_{s}^{2}\left\{ T^{A},T^{B}\right\} \gamma_{3}\delta_{\mu3}\delta_{\nu3}e^{-ip_{3}z}\left(\frac{z}{i\widehat{q}_{3}}e^{\frac{z}{\left|z\right|}i\frac{aq_{3}}{2}}-\frac{a\left|z\right|}{2}\right)\ .
\end{equation}
This terms, however, is proportional to $\delta'(x-1)$ after the Fourier transformation (with respect to $z$) to $x$ dependence and it is also excluded in the calculation since $\delta'(x-1)$ only contributes at $x=1$, similar to the virtual corrections.

The fermion propagator, gluon propagator and quark-gluon-quark vertex
can be rewritten in short form for future convenience
\bseq\label{eq:FynmnRl_WC}
\begin{align}
S_{F}\left(k\right)=\frac{-i\mathcal{K}\!\!\!\!/+\mathcal{M}}{\mathcal{K}^{2}+\mathcal{M}^{2}},
\end{align}
\begin{align}
D_{g,\mu\nu}\left(P-k\right)=\frac{\delta_{\mu\nu}}{\mathcal{Q}^{2}},
\end{align}
\begin{align}
V_{\alpha}^{a}\left(P,k,P-k\right)= & -ig_{s}T^{a}\varLambda_{\alpha}\gamma_{\alpha}-ig_{s}T^{a}
\varOmega_{\alpha}-ig_{s}T^{a}\sum_\rho\sigma_{\alpha\rho}\widehat{2\left(P-k\right)}_{\rho}\varXi_{\alpha}.
\end{align}
\eseq where we have defined
\bseq\label{eq:FynmnRl_WC_dfntn}
\begin{align}
&\mathcal{K}_{\mu}=\widehat{2k}_{\mu},\hspace{-6em}&&\mathcal{M}=r\sum_{\mu}\left(\frac{2}{a}-\widetilde{2k}_{\mu}\right)+2m,\\
&\mathcal{P_{\mu}=}\widehat{2P}_{\mu},\hspace{-6em}&&\varDelta=r\sum_{\mu}\left(\frac{2}{a}-\widetilde{2P}_{\mu}\right)+2m,\label{eq:calP_varDlt}\\
&\mathcal{Q}_{\mu}=\widehat{k-P}_{\mu},\hspace{-6em}&&\varLambda_{\mu}=\frac{a}{2}\widetilde{k+P}_{\mu},\\
&\varOmega_{\mu}=\frac{ar}{2}\widehat{k+P}_{\mu},\hspace{-6em}&& \varXi_{\mu}=r\frac{c_{\mathrm{SW}}a^{2}}{8}\widetilde{k-P}_{\mu}.
\end{align}
\eseq 

\emphz{The Feynman rules for na\"ive lattice fermion action corresponds to $r=0$ limit of Eq.~(\ref{eq:FynmnRl_WC},\ref{eq:FynmnRl_WC_dfntn})}
\bseq
\begin{align}
\varDelta^{(0)}&=\mathcal{M}^{(0)}=2m,\\
  V^{(0),a}_\alpha&=-igT^a\varLambda_\alpha\gamma_\alpha.
\end{align}
\eseq

\emphs{We now perform calculations of the one-loop quark quasi-PDF with both the na\"ive lattice fermion action and Wilson-Clover fermion action, and discuss the impact of the Wilson-Clover term.}

\subsection{Quasi-PDF in na\"ive lattice fermion action}

In the LPT quasi-PDF calculation, we use the same
configuration as in Ref.~\cite{Xiong:2013bka}: the momentum of the parent
quark is set to $P_{\mu}=\left(0,0,P_{3},P_{4}\right)$ and the transverse
momentum UV cut-off is naturally chosen to be the lattice momentum
cut-off $\pm\pi/a$. \emphs{The relevant Feynman rules in na\"ive lattice fermion action reads:}
\bseq
\begin{align}\label{eq:Oa0_prpgtr}
&S_F^{(0)}\left(k\right)=\frac{i\mathcal{K}\!\!\!\!/+\mathcal{M}^{0}}{\mathcal{K}^2+\left(\mathcal{M}^{0}\right)^2},
&&D_{g,\mu\nu}^{(0)}\left(P-k\right)=\frac{\delta_{\mu\nu}}{\mathcal{Q}^2},\\
&V_{\alpha}^{(0),a}\left(P,k,P-k\right)= -ig_{s}T^{a}\varLambda_{\alpha}\gamma_{\alpha},
&&\left[\sum_sU_s\left(P\right)\overline{U}_s\left(P\right)\right]^{(0)}= i\mathcal{P}\!\!\!\!/+\varDelta^{(0)}.
\end{align}
\eseq

After some algebra, the one-loop correction
diagrams give \bseq
\begin{align}
\tilde{q}_{a}^{\mathrm{nv}}\!\left(x\right)\!= \!& \int_{-\frac{\pi}{a}}^{\frac{\pi}{a}}\frac{d^{4}k}{\left(2\pi\right)^{4}}\,\left(-4g_{s}^{2}C_{F}\right)\!P_{3}\!
\left\{\!\frac{2\mathcal{K}_{3}\!\left[\mathcal{P}_{3}\mathcal{K}_{3}\!\left(\varLambda^{2}\!-\!2\varLambda_{3}^{2}\right)
\!+\!\varDelta^{(0)}\mathcal{M}^{(0)}\varLambda^{2}\right]\!\!+\!2\mathcal{P}_{4}\mathcal{K}_{3}
\mathcal{K}_{4}\!\left(\varLambda^{2}\!-\!2\varLambda_{4}^{2}\right)}{\mathcal{P}_{3}\mathcal{Q}^{2}
\left[\mathcal{K}^{2}\!+\!\left(\mathcal{M}^{(0)}\right)^{2}\right]^{2}}\right.\nonumber \\
 & \left.-\frac{\left(\varLambda^{2}-2\varLambda_{3}^{2}\right)}{\mathcal{Q}^{2}\left[\mathcal{K}^{2}+\left(\mathcal{M}^{(0)}\right)^{2}\right]}\right\} \delta\left(k_{3}-xP_{3}\right),
\end{align}
\begin{align}
\tilde{q}_{b}^{\mathrm{nv}}\left(x\right)=\tilde{q}_{c}^{\mathrm{nv}}\left(x\right)= & 2g_{s}^{2}C_{F}\int_{-\frac{\pi}{a}}^{\frac{\pi}{a}}\frac{d^{4}k}{\left(2\pi\right)^{4}}\,\frac{P_{3}\varLambda_{3}}{\mathcal{P}_{3}\mathcal{Q}_{3}}\frac{\mathcal{K}_{4}\mathcal{P}_{4}-\mathcal{K}_{3}\mathcal{P}_{3}+\varDelta^{(0)}\mathcal{M}^{(0)}}{\mathcal{Q}^{2}\left[\mathcal{K}^{2}+\left(\mathcal{M}^{(0)}\right)^{2}\right]}\delta\left(k_{3}-xP_{3}\right),
\end{align}
\begin{align}\label{eq:q0d_k_intgrnd}
\tilde{q}_{d}^{\mathrm{nv}}\left(x\right)= & -g_{s}^{2}C_{F}\int_{-\frac{\pi}{a}}^{\frac{\pi}{a}}\frac{d^{4}k}{\left(2\pi\right)^{4}}\,\frac{P_{3}}{\mathcal{Q}_{3}^{2}\mathcal{Q}^{2}}\delta\left(k_{3}-xP_{3}\right).
\end{align}
\eseq 
The superscript nv denotes the quasi-PDF calculated in na\"ive fermion action. In the above equations, those terms odd under $k_{1,2}\rightarrow-k_{1,2}$
will not contribute to the integral and they are therefore already omitted.

In order to analytically integrate out $k_{4}$, we introduce a variable
change $z=a^{-2}e^{iak_{4}}$~\cite{Muller:2009af,C.J.Monahan}, then the $k_{4}$-integration is transformed
to a contour integral along the circle on the complex plane
\begin{equation}
\int_{-\frac{\pi}{a}}^{\frac{\pi}{a}}dk_{4}\,f\left(k_{4}\right)=\frac{-i}{a}\ointctrclockwise_{\left|z\right|=a^{-2}}\frac{dz}{z}\,f\left(\frac{-i}{a}\ln\left(a^{2}z\right)\right).
\end{equation}
Under this transformation, the denominators of the quark and gluon propagators
can be rewritten as 
\bseq
\begin{align}\label{eq:r0_z_pole}
\mathcal{D}_{F}^{(0)}=\mathcal{K}^{2}+\left(\mathcal{M}^{(0)}\right)^{2}= & -a^{-2}z^{-2}\left(a^{2}z^{2}-\Gamma_{+}\right)\left(a^{2}z^{2}-\Gamma_{-}\right),
\end{align}
\begin{equation}
\mathcal{D}_{g}^{(0)}=\mathcal{Q}^{2}=-e^{-iaP_{4}}z^{-1}\left(z-\Pi_{-}\right)\left(z-\Pi_{+}\right).
\end{equation}
\eseq where $\Gamma_{\pm}$ and $\Pi_{\pm}$ are defined as
\begin{align}\label{eq:z_poles}
\Gamma_{\pm}=\frac{\kappa\pm\sqrt{\kappa^{2}-\frac{4}{a^{4}}}}{2},\; & \Pi_{\pm}=e^{iaP_{4}}\frac{\eta\pm\sqrt{\eta^{2}-\frac{4}{a^{4}}}}{2},
\end{align}
and
\begin{align}
\kappa=\sum_{j=1}^{3}\mathcal{K}_{j}^{2}+\left(\mathcal{M}^{(0)}\right)^{2}+\frac{2}{a^{2}},\; & \eta=\sum_{j=1}^{3}\mathcal{Q}_{j}^{2}+\frac{2}{a^{2}}.
\end{align}
The $z$-poles inside the integration circle $\left|z\right|=a^{-2}$
are $\pm\sqrt{\Gamma_{-}}/a$ and $\Pi_{-}$. The corresponding $k_{4}=-ik^{0}$
poles reduce to the continuum $k^{0}$ poles on the upper complex
plane in the $a\rightarrow0$ limit \bseq
\begin{align}
&{k}^{g\hphantom{,+}}_{4}= -\frac{i}{a}\log\left(\!a^{2}\Pi_{-}\!\right)\hspace{-3em}&&\rightarrow\;\; k^0_{g\hphantom{,+}} = P^{0}-\sqrt{\boldsymbol{k}_{\perp}^{2}+\left(k_{3}-P_{3}\right)^{2}+m^{2}-i\epsilon},\\
&{k}^{q,+}_{4}= -\frac{i}{a}\log\left(a\sqrt{\Gamma_{-}}\right)\hspace{-3em}&&\rightarrow\;\; k^0_{q,+}= -\sqrt{\boldsymbol{k}_{\perp}^{2}+k_{3}^{2}+m^{2}-i\epsilon},\\
&{k}^{q,-}_{4}= -\frac{i}{a}\log\left(\!-a\sqrt{\Gamma_{-}}\!\right)\hspace{-3em}&&\rightarrow\;\; k^0_{q,-}=\frac{i\pi}{a}-\sqrt{\boldsymbol{k}_{\perp}^{2}+k_{3}^{2}+m^{2}-i\epsilon}~,
\end{align}
\eseq except for the second quark pole ${k}^{q,-}_{4}$, which turns out
to present an unphysical continuum limit. However, the residue at
this unphysical quark pole $\bar{k}^{q,-}_{4}$ vanishes in the continuum
limit and the unphysical pole decouples in continuum limit. Applying
the above transformation, the integral over $k_{4}$ is equivalent
to taking the residue of the transformed integrand at $\bar{k}^{q,\pm}_{0}$,
$\bar{k}^{g}_{4}$. 

{We have performed a Wick rotation $P_4\rightarrow-iP^0$ ($P^0$ takes the value of r.h.s of Eq.~\eqref{eq:OnShell}) in order to compare the quasi-PDF in LPT with the continuum quasi-PDF, which is calculated with Minkowski parent quark momentum $P$ and loop momentum $k$. The impact of the Wick rotation will be discussed in a forthcoming paper~\cite{Ji:2017rah}.}

{It should be noticed that, in this work, the Wick rotation should take place after $k_4$ having been integrated out. One can see from the $z$ poles in Eq.{~\eqref{eq:z_poles}}, if the Wick rotation is performed before taking residue at poles inside the integration circle $z=a^{-2}$, the gluon poles in Eq.~\eqref{eq:z_poles} become
\begin{equation}\label{eq:Wick_Rt_g_pole} 
\frac{e^{iaP^4}}{2}\left(\eta\pm\sqrt{\eta^2-\frac{4}{a^4}}\right)\rightarrow \frac{e^{aP^0}}{2}\left(\eta\pm\sqrt{\eta^2-\frac{4}{a^4}}\right).
\end{equation}
In this case, one can not ensure that the gluon poles are inside the integration circle due to the exponential factor $e^{aP^0}$ being larger than one, thus as a consequence, the residue at the gluon pole may not be included in the integration and it leads to incorrect results. The kinematic region which constrains the Wick-rotated gluon pole ($\Pi_-$ with $P_4\rightarrow -iP^0$) inside the integration circle $z=a^{-2}$ can be evaluated by
\begin{equation}\label{eq:kt_cnstrn_LPT}
\frac{e^{aP^0}}{2}\left(\eta-\sqrt{\eta^2-\frac{4}{a^4}}\right)<\frac{1}{a^2}.
\end{equation}
Expanding to $\mathcal{O}\left(a^1\right)$  gives
\begin{equation}\label{eq:kt_cnstrn}
1+a\left(\sqrt{m^2+P_3^2}-\sqrt{\boldsymbol{k}_\perp^2+P_3^2\left(1-x\right)^2}\right)<1\Rightarrow \boldsymbol{k}_\perp^2>m^2+x(2-x)P_3^2.
\end{equation}}
For the region $\boldsymbol{k}_\perp^2<m^2+x(2-x)P_3^2$, as discussed in Ref.~\cite{Briceno:2017cpo}, one needs to consider another piece of integration to recover the correct integral. It is due to the gluon pole crossing the integration circle. The condition in Eq.~\eqref{eq:kt_cnstrn}, which determines the position of gluon pole, is only an $\mathcal{O}\left(a^1\right)$ approximation to the exact condition (Eq.\eqref{eq:kt_cnstrn_LPT})in LPT. The exact condition in LPT is hard to solve,
thus in our quasi-PDF calculation, we apply the Wick rotation $P_4\rightarrow -iP^0$ after $k_4$ being integrated out to avoid 
the complexity.

The following contents of this paper are based on the $k_4$-integrated out results (except Sec.~\ref{sec:Oa1}, the quasi-PDF calculated with Wilson-Clover action), that means that we have already performed the Wick rotation $P_4\rightarrow-iP^0$ after the $k_4$ integration unless specified.

The expressions of the $k_4$ integration in Eq.~(\ref{eq:f0_abcd}) are very lengthy and they do not provide any insight to the discussion here. We therefore provide these expressions in appendix~\ref{Oa0_k4_intgrtn_exprsn}, see Eqs.~(\ref{eq:fa0k4Inted},\ref{eq:fbc0k4Inted},\ref{eq:fd0k4Inted}).

\subsection{Continuum limit and collinear behavior of quasi-PDF in na\"ive lattice fermion action}\label{Sec:Cntnm_ClnrDvgnc}
The continuum limit of Eqs.~(\ref{eq:fa0k4Inted},\ref{eq:fbc0k4Inted},\ref{eq:fd0k4Inted})
can be calculated directly by performing the Wick rotation $P_{4}\rightarrow-iP^0$
and the $a\rightarrow0$ limit. The result turns out to be identical to
the $\boldsymbol{k}_{\perp}$-unintegrated quasi-PDF calculated directly
in the continuum. We will show the equivalence later in the example calculation of $\tilde{q}^{\mathrm{nv}}_b\left(x\right)$.

There is a major difference between the quasi-PDF calculated by LPT and the
continuum quasi-PDF: The collinear divergence is absent in the LPT-calculated quasi-PDF when the lattice spacing $a$ is finite, even after performing the Wick rotation $P_4\rightarrow -iP^0$. 

The reason is that the lattice artifacts from the finite lattice spacing have regulated the
collinear divergence. The absence of collinear divergence can be verified
in the numerical results by setting the quark mass to zero, see Fig.~\ref{fg:f_0m}.  It can  also be seen in the analytical $\boldsymbol{k}_{\perp}$-integrand
of  the quasi-PDF in LPT. We take the calculation of diagram ${b}$ as
an example to present the continuum limit and illustrate how to extract the collinear behavior in lattice perturbation calculation.

The continuum correspondence of $\tilde{q}^{\mathrm{nv}}_b\left(x\right)$ is
\begin{align}
\left[\tilde{q}_b\left(x\right)\right]_\mathrm{cont.}= \int d^4k \frac{g_s^2C_F}{\left(2\pi\right)^4} \frac{\bar{U}\left(P\right)\gamma^z\left(k\!\!\!/+m\right)\gamma^zU\left(P\right)}
{\left(P^z-k^z\right)\left(P-k\right)^2\left(k^2-m^2\right)}\delta\left(x-\frac{k^z}{P^z}\right)/\left[\bar U\left(P\right)\gamma^z U\left(P\right)\right]
\end{align}
The $k^0,k^z$-integration gives exactly the same result as $a\rightarrow 0$ limit of the lattice perturbation result \eqref{eq:fbc0k4Inted}
\begin{align}
\notag &\lim_{a\rightarrow 0}\tilde{q}^{\mathrm{nv}}_b\left(x,\boldsymbol{k}_\perp\right) \\
\notag=&\frac{\alpha_s C_F}{8\pi^3P_3\left(1-x\right)} \left[-\frac{P_0 \sqrt{\boldsymbol{k}_{\perp}^2+m^2+P_3^2 x^2}+m^2-P_3^2 x}{\sqrt{\boldsymbol{k}_{\perp}^2+m^2+P_3^2 x^2}P_0 \sqrt{\boldsymbol{k}_{\perp}^2+m^2+P_3^2 x^2}+m^2+P_3^2 x}\right.\\
&\left.+\frac{\left(P_0 \left(\sqrt{\boldsymbol{k}_{\perp}^2\!-\!m^2\!+\!P_3^2 (x\!-\!2) x\!+\!P_0^2}\!-\!P_0\right)\!+\!m^2\!-\!P_3^2 x\right)}{\sqrt{k_{\perp}^2\!-\!m^2\!+\!P_3^2\! (x\!-\!2) x\!+\!P_0^2}\! \left(\!P_0\! \left(\!\sqrt{\boldsymbol{k}_{\perp}^2\!-\!m^2\!+\!P_3^2 (x\!-\!2) x\!+\!P_0^2}\!-\!P_0\!\right)\!+\!m^2\!+\!P_3^2 x\right)}\!\right]~.
\end{align}
The continuum quasi-PDF can be obtained by integrating out $\boldsymbol {k}_\perp$,
where we already performed the expansion $m\rightarrow 0$. It clearly shows that there exists a collinear divergence in the region $0<x<1$
\begin{equation}\label{eq:qb_cont_col}
\left[\tilde q_b\left(x\right)\right]_{cont.,col.}=
\begin{cases}
-\frac{g_s^2 C_F}{8\pi^2}\frac{2x}{1-x}\ln m^2+\cdots & 0<x<1\\
\cdots & x<0 \text{ or } x>1
\end{cases}
\end{equation}
where the ellipsis denotes those terms which do not contain a collinear divergence (at the order of $\mathcal{O}\left(m^0\right)$).

Physically, the collinear divergence happens when the split quark's momentum $\boldsymbol{k}$ is parallel to the
parent quark's momentum $\boldsymbol{P}$, or equivalently $\boldsymbol{k}_\perp=\boldsymbol{0}$. To extract the collinear behavior of
the LPT-calculated quasi-PDF, we expand the numerator and denominator of Eq.~\eqref{eq:fbc0k4Inted} around
$\left|\boldsymbol{k}_\perp\right|=0$. The expansion takes the form
\begin{equation}
\lim_{k_\perp\rightarrow \boldsymbol{0}}\tilde{q}^{\mathrm{nv}}\left(x,\boldsymbol{k}_\perp\right)=\sum_{i=1}^{3}\frac{\mathcal{N}_{b,i}^{(0)}}
{\mathcal{D}_{b,i}^{(0)}+\mathcal{D}_{b,i}^{(1)}\boldsymbol{k}_\perp^2},
\end{equation}
in which $i=1,2,3$ denotes the 1st, 2nd, 3rd term in \eqref{eq:fbc0k4Inted} corresponding to the gluon pole, quark pole and unphysical quark pole's residue. The expressions for $\mathcal{N}_{i,b}^{(n)}$ and $\mathcal{D}_{i,b}^{(n)}$ are
\bseq
\begin{align}
\mathcal{N}_{b,1}^{(0)}=&	ia^{3}P_{3}\varDelta_{\Pi}g_{s}^{2}C_F\frac{\cos\left(\frac{aP_{3}}{2}(x\!+\!1)\right)}{\sin\left(\frac{aP_{3}}{2}(x-1)\right)}
\left[\!\frac{2ia^{2}m^{2}\varDelta_{\Pi}\!+\!\left(\varDelta_{\Pi}^{2}\!-\!1\right)\sin\left(aP_{4}\right)}{\sin\left(aP_{3}\right)}
\!-\!\frac{2i\varDelta_{\Pi}}{\csc\left(aP_{3}x\right)}\vphantom{\frac{\varDelta_{\Pi}^{2}}{P_{3}}}\right],\\
\mathcal{D}_{b,1}^{(0)}=&16 \pi ^3 \sqrt{\mathcal{R}_{\Pi }^2-1} \left(\Delta _{\Pi }^4-2 \Delta _{\Pi }^2 \mathcal{R}_{\Gamma }+1\right),\\
\mathcal{D}_{b,1}^{(1)}=&32 \pi ^3 a^2 \Delta _{\Pi }^2 \left(\mathcal{R}_{\Gamma }-\Delta _{\Pi }^2\right)+\frac{8 \pi ^3 a^2 \left(8 \Delta _{\Pi }^2\left(1-\mathcal{R}_{\Pi }^2\right)+\mathcal{R}_{\Pi } \left(\Delta _{\Pi }^4-2 \Delta _{\Pi }^2 \mathcal{R}_{\Gamma }+1\right)\right)}{\sqrt{\mathcal{R}_{\Pi }^2-1}},
\end{align}
\eseq
\bseq
\begin{align}
\notag \mathcal{N}_{b,2}^{(0)}=&	a^{3}P_{3}e^{iaP_{4}}g_{s}^{2}C_{F}\frac{\cos\left(\frac{aP_{3}}{2}(x+1)\right)}{\sin\left(\frac{aP_{3}}{2}(x-1)\right)}
\left[-\frac{2a^{2}m^{2}\sqrt{\varDelta_{\Gamma}}+i\left(1-\varDelta_{\Gamma}\right)\sin\left(aP_{4}\right)}
{\sin\left(aP_{3}\right)}\right.\\	&\left.+2\sqrt{\varDelta_{\Gamma}}\sin\left(aP_{3}x\right)\vphantom{\frac{\sqrt{\varDelta_{\Gamma}}}{P_{3}}}\right],\\
\mathcal{D}_{b,2}^{(0)}=&32\pi^{3}\sqrt{\mathcal{R}_{\Gamma}^{2}-1}\left(-2e^{iaP_{4}}\sqrt{\Delta_{\Gamma}}
\mathcal{R}_{\Pi}+e^{2iaP_{4}}+\Delta_{\Gamma}\right),\\
\notag \mathcal{D}_{b,2}^{(1)}=&\frac{32\pi^{3}a^{2}}{\sqrt{\mathcal{R}_{\Gamma}^{2}-1}}\left[2e^{iaP_{4}}
\sqrt{\varDelta_{\Gamma}}\sqrt{\mathcal{R}_{\Gamma}^{2}-1}\mathcal{R}_{\Pi}-e^{iaP_{4}}\sqrt{\varDelta_{\Gamma}}\left(4\mathcal{R}_{\Gamma}\mathcal{R}_{\Pi}
+\mathcal{R}_{\Gamma}^{2}-1\right)\right.\\
&\left.+2e^{2iaP_{4}}\mathcal{R}_{\Gamma}+2\varDelta_{\Gamma}^{2}
\vphantom{\sqrt{\mathcal{R}_{\Gamma}^{2}}}\right].
\end{align}
\eseq
Here, we have omitted the unphyscial quark pole ($i=3$) because it is of $\mathcal{O}\left(a^2\right)$ in the lattice spacing and does not contain any collinear divergence. The definitions of $\mathcal{R}_{\Gamma,\Pi}$ and $\varDelta_{\Gamma,\Pi}$
are \bseq
\begin{align}
&\mathcal{R}_{\Pi}=\left.\frac{1}{2}\left(\sqrt{\frac{\Pi_{-}}{\Pi_{+}}}+\sqrt{\frac{\Pi_{+}}{\Pi_{-}}}\right)\right|_{\boldsymbol{k}_{\perp}=\boldsymbol{0}},
&&\mathcal{R}_{\Gamma}=\left.\frac{1}{2}\left(\sqrt{\frac{\Gamma_{-}}{\Gamma_{+}}}+\sqrt{\frac{\Gamma_{+}}{\Gamma_{-}}}\right)\right|_{\boldsymbol{k}_{\perp}=\boldsymbol{0}},\\
&\varDelta_{\Pi}=e^{iaP_{4}}\left(\mathcal{R}_{\Pi}-\sqrt{\mathcal{R}_{\Pi}^{2}-1}\right),
&&\varDelta_{\Gamma}=\mathcal{R}_{\Gamma}-\sqrt{\mathcal{R}_{\Gamma}^{2}-1}.
\end{align}
\eseq
The $\boldsymbol{k}_\perp$-integration of $\mathcal{N}_{b,i}^{(0)}/\left(\mathcal{D}_{b,i}^{(0)}+\mathcal{D}_{b,i}^{(1)}\boldsymbol{k}_\perp^2\right)$ leads to
\begin{equation}
\int_0^\mu d^2\boldsymbol{k}_\perp \frac{\mathcal{N}_{b,i}^{(0)}}{\left(\mathcal{D}_{b,i}^{(0)}+\mathcal{D}_{b}^{(1)}\boldsymbol{k}_\perp^2\right)}
=\pi\frac{\mathcal{N}_{b,i}^{(0)}}{\mathcal{D}_{b,i}^{(1)}}\ln\frac{\mu^2\mathcal{D}_{b,i}^{(1)}}{\mathcal{D}_{b,i}^{(0)}}~,
\end{equation}
where $\mu$ is a finite scale which does not affect the IR behavior of the integral. It straightforward to check that $\mathcal{D}_{b,i}^{(0)}$ is nonzero when $m\rightarrow 0$ with finite lattice spacing $a$.
Consequently, the collinear divergence is regularized by $\mathcal{D}_{b,i}^{(0)}$. We have calculated the quasi-PDF numerically with explicit $m=0$ as shown in Fig.\ref{fg:f_0m} and there is no
signal of a collinear divergence in the whole $x$ region. If we take the continuum limit before the expansion around $m=0$, we have
\begin{equation}\label{eq:qbc_clnr_1}
\lim_{m\rightarrow 0}\left(\lim_{a\rightarrow 0 }\pi\frac{\mathcal{N}_{b,1}^{(0)}}{\mathcal{D}_{b,1}^{(1)}}
\ln\frac{\mu^2\mathcal{D}_{b,1}^{(1)}}{\mathcal{D}_{b,1}^{(0)}}\right)=
\begin{cases}
-\frac{g_s^2 C_F}{8\pi^2}\frac{2x}{1-x}\ln m^2+\cdots & x<1\\
\cdots & {x>1}
\end{cases}
\end{equation}
Through the above procedure, we can extract the collinear behavior of the 2nd term in \eqref{eq:fbc0k4Inted}, which reads
\begin{equation}\label{eq:qbc_clnr_2}
\lim_{m\rightarrow 0}\left(\lim_{a\rightarrow 0}\pi\frac{\mathcal{N}_{b,2}^{(0)}}{\mathcal{D}_{b,2}^{(1)}}
\ln\frac{\mu^2\mathcal{D}_{b,2}^{(1)}}{\mathcal{D}_{b,2}^{(0)}}\right)=
\begin{cases}
\frac{g_s^2 C_F}{8\pi^2}\frac{2x}{1-x}\ln m^2+\cdots & x<0\\
\cdots & x>0
\end{cases}
\end{equation}
To sum up, the collinear divergence of $\tilde{q}_b\left(x\right)$ is
\begin{equation}\label{eq:qbc_clnr_tot}
\sum_{i=1}^2\lim_{m\rightarrow 0}\left(\lim_{a\rightarrow 0}\pi\frac{\mathcal{N}_{b,i}^{(0)}}{\mathcal{D}_{b,i}^{(0)}}
\ln\frac{\mu^2\mathcal{D}_{b,i}^{(1)}}{\mathcal{D}_{b,i}^{(0)}}\right)=
\begin{cases}
\emphz{-}\frac{g_s^2 C_F}{8\pi^2}\frac{2x}{1-x}\ln m^2+\cdots & 0<x<1\\
\cdots & x<0 \text{ or } x>1
\end{cases}
\end{equation}
and it is identical to the collinear divergence of the continuum quasi-PDF in \eqref{eq:qb_cont_col}. The above example calculation has shown that in order to recover the collinear divergence in continuum quasi-PDF,
one needs to eliminate the effect of the finite lattice spacing by taking $a\rightarrow 0$ limit before the expansion around $m=0$. {Eqs.~(\ref{eq:qbc_clnr_1},\ref{eq:qbc_clnr_2},\ref{eq:qbc_clnr_tot}) show that the collinear divergence originates from the gluon pole $i=1$ in the region 
$0<x<1$.  We further expand $\mathcal{D}_{b,1}^{(0)}/\mathcal{D}_{b,1}^{(1)}$ to 
$\mathcal{O}\left(a^2\right)$ (in the region $0<x<1$, there is no collinear divergence in region $x<0\lor x>1$) in order to study how lattice artifacts affect the 
collinear behavior
\begin{align}
\lim_{a\rightarrow 0 }\ln \frac{\mu^2\mathcal{D}_{b,1}^{(0)}}{\mathcal{D}_{b,1}^{(0)}} 
\overset{\scriptsize{P_3\gg m}}{\approx} \ln \frac{\mu^2}{\frac{1}{2} a^2P_3^4 \left[x \left(x^2-1\right)^2 + \mathcal{O}\left(\frac{m^2}{P_3^2}\right)\right]+m^2 (x-1)^2}.
\end{align}}
Therefore, the correct limit to recover the logarithmic collinear divergence is $aP_3^2\approx m$ and 
$m\ll P^3$. The first condition $aP^3_2\approx m$ matches the lattice artifacts with the collinear regulator $m^2$, since,
assuming $aP_3^2=\lambda m$ with $\lambda$ being a finite constant, gives
\begin{align}
\ln \frac{\mu^2}{\frac{1}{2} a^2P_3^4 \left[x \left(x^2-1\right)^2\right]+m^2 (x-1)^2}=\ln \frac{\mu^2}{m^2}-\ln\left[(x-1)^2+x(x^2-1)^2\frac{\lambda^2}{2}\right]
\end{align}
and the collinear divergent part $\ln \left(\mu^2/m^2\right)$ stays the same as long as $aP_3^2$ and $m$ are of the same order. This condition also indicates $aP_3\ll 1$ and it eliminates the subleading order lattice artifacts (e.g. $a^3 m^2P_3^3$, $a^4P_3^6$).
The second condition $m\ll P_3$ justifies $m$ as a collinear regulator. The limit to fully recover the quasi-PDF calculated in continuum QCD is $aP_3^2\ll m \ll P_3$, in which the first condition allows us to drop the lattice artifacts completely
\begin{align}
\ln \frac{\mu^2}{\frac{1}{2} a^2P_3^4 \left[x \left(x^2-1\right)^2\right]+m^2 (x-1)^2}\overset{aP_3^2\ll m \ll P_3}{\longrightarrow}\ln \frac{\mu^2}{m^2(1-x)^2}.
\end{align}
We can also see that the effects 
of lattice artifacts are getting enhanced for large $P_3$. We will further justify 
this conclusion in the numerical calculations, see Sec.\ref{sctn:nmrc_rslt}.

\subsection{One-loop quasi-PDF in Wilson-Clover lattice fermion action}\label{sec:Oa1}

The calculation ($k_{4}$-integration) of the quasi-PDF in Wilson-Clover action is analogous to the na\"ive lattice fermion calculation, however, the analytical $k_{4}$-integration in Wilson-Clover lattice fermion case are too cumbersome to be displayed (and again do not provide any insight to our present discussion), hence we only briefly introduce the procedure of $k_4$-integration. {(we do not apply the Wick rotation $P_4\rightarrow-iP^0$ in this subsection, because $k_4$ has not  been integrated out)}.

\emphs{In the case of Wilson-Clover lattice fermion action, there are in total four poles of $z=a^{-2}e^{iak_4}$ in the quark propagator ~\eqref{eq:S_F}:
\begin{align}
\notag  z_{\lambda_1,\lambda_2}=&\frac{1}{2}\left\{\frac{r\left(-2am\!-\!8r\!+\!ar\sum_{i=1}^3\widetilde{2k}_i\right)\!+\!\lambda_2\sqrt{\Phi}}{a^2\left(1-r^2\right)}
\!+\!\lambda_1\left[-\frac{2}{a^4}\!+\!\frac{2r\left(\!-\!2am\!-\!8r\!+\!ar\sum_{i=1}^3\widetilde{2k}_i\right)}{a^4\left(1-r^2\right)^2}\right.\right.\\
&\left.\left.\times\left(\!r\left(\!-\!2am\!-\!8r\!+\!ar\sum_{i=1}^3\widetilde{2k}_i\right)+\lambda_2\sqrt{\Phi}\right)+\frac{\Phi+a^2r^2\sum_{i=1}^3\widehat{2k}_i^2\!+\!6r^2\!-\!2}{a^4\left(1-r^2\right)}\right]^{\tfrac{1}{2}}\right\},
\end{align}
where $\lambda_1=\pm 1$ and $\lambda_1=\pm 1$ and
\begin{align}
 \notag \Phi=&4+4a^2m^2+32amr+60r^2+a^2r^2{\widetilde{2\boldsymbol{k}}}^2_\perp -4ar\left(am+4r\right)\widetilde{2k}_3+a^2r^2\widetilde{2k}_3^2\\
  &+2ar\sum_{i=1}^2{\widetilde{2k}_i}\left(-2am-8r+ar\widetilde{2k}_3\right)+2a^2r^2\widetilde{2k}_1\widetilde{2k}_2+a^2\left(1-r^2\right){\widetilde{\boldsymbol{2k}}}^2.
\end{align}
where $\widetilde{2\boldsymbol{k}}_\perp^2=\sum_{i=1}^2\widetilde{2\boldsymbol{k}}^2_i$ and $\widetilde{2\boldsymbol{k}}^2=\sum_{i=1}^3\widetilde{2\boldsymbol{k}}^2_i$. In the $r\rightarrow 0$ limit, $|z_{\pm 1,\pm 1}|$ become larger than $a^{-2}$ and are therefore excluded from the integration, while $z_{\pm 1,\mp 1}$ reduce to the $z$ pole in Eq.(\ref{eq:r0_z_pole}), i.e. the na\"ive fermion action result. The $z$ poles of gluon propagator: $\Pi_\pm$ in eq.~(\ref{eq:z_poles}) are unchanged.}

\emphz{For $r>0$, the $z$-poles inside the integration circle $|z|=a^{-2}$ are $z_{\mp 1,\pm 1}$ and $z=\Pi_-$. Taking the residue of eq.~(\ref{eq:f0_abcd}) at the above mentioned poles gives the $k_4$-integration of one-loop quasi-PDF in Wilson-Clover lattice fermion action.}

\emphz{It should be noted that after the $k_4$ integration, $\tilde{q}_d(x)$ is independent of $r$ and quark mass $m$.  Therefore this quantity is the same for both the na\"ive fermion action and Wilson-Clover fermion action cases and we only present the numerical results of $\tilde{q}_d(x)$ with na\"ive fermion action.}

\section{numerical results }\label{sctn:nmrc_rslt}

{After analytically integrating out $k_{3}$, $k_{4}$ and applying the Wick rotation $P_4\rightarrow-iP^0$} to Eqs.~(\ref{eq:fa0k4Inted},\ref{eq:fbc0k4Inted},\ref{eq:fd0k4Inted}), the transverse
momentum $\boldsymbol{k}_{\perp}$ integration is performed through
a two-dimensional numerical integration. The $k$-integrated quasi-PDF
actually only depends on three dimensionless quantities:
$x$, $am$ and $aP_{3}$. However, in order to explore how the quasi-PDF calculated by lattice
perturbation theory evolves with respect to $a$, $m$ and $P_{3}$,
we still keep them as independent parameters. The lattice spacing
is chosen as $a=2^{n}~\mathrm{fm}$ for $n=-11,-10,-8,-4,-3,-2,-1$, the parent quark's longitudinal momentum is $P_{3}=1.5\pi\,\text{fm}^{-1}$ and
bare quark mass is set to $m=0.5\pi\,\text{fm}^{-1}$ or $m=0.005\pi\,\text{fm}^{-1}$ to explore the
collinear behavior of the quasi-PDF. We also calculated the numerical
results of quasi-PDF in LPT for the vanishing bare quark mass case {together with  
bare quark mass $m$ set to the one-loop result for the critical mass of Wilson-Clover fermions,} in order to confirm the absence of a collinear divergence. We have omitted the quasi-PDF in the region $0.6<x<1.4$ in order
to avoid the divergent terms such as $(1-x)^{-1}$ and $(1-x)^{-2}$
which will become overwhelming compared to other $x$ regions. We
separate the contribution from diagrams $a$, $b$, $c$, $\tilde{q}_{a,b,c}\left(x\right)$,
and diagram $d$, $\tilde{q}_{d}\left(x\right)$, because $\tilde{q}_{d}\left(x\right)$
is linearly UV divergent and also contains a quadratic pole,
i.e. $\left(1-x\right)^{-2}$. The $\tilde{q}_{d}\left(x\right)$ contribution
to the quasi-PDF is much larger than $\tilde{q}_{a,b,c}\left(x\right)$.
In such a case, the information of $\tilde{q}_{a,b,c}\left(x\right)$
will be overwhelmed by $\tilde{q}_{d}\left(x\right)$ if the two parts
are summed together. Furthermore, the collinear behavior of $\tilde{q}_{a,b,c}\left(x\right)$
and $\tilde{q}_{d}\left(x\right)$ is also quite different in the
continuum quasi-PDF: the former one carries collinear divergence in
the region $0<x<1$ while the latter one is free of collinear divergence.
For those reasons, we present the numerical results of $\tilde{q}_{a,b,c}\left(x\right)$
and $\tilde{q}_{d}\left(x\right)$ separately. For the numerical calculation with Wilson-Clover fermion, the Wilson parameter
is set to $r=1/2$ and $c_{\mathrm{SW}}=1$ for the leading order Clover parameter.

\begin{figure}
\begin{centering}
\includegraphics[width=1\textwidth]{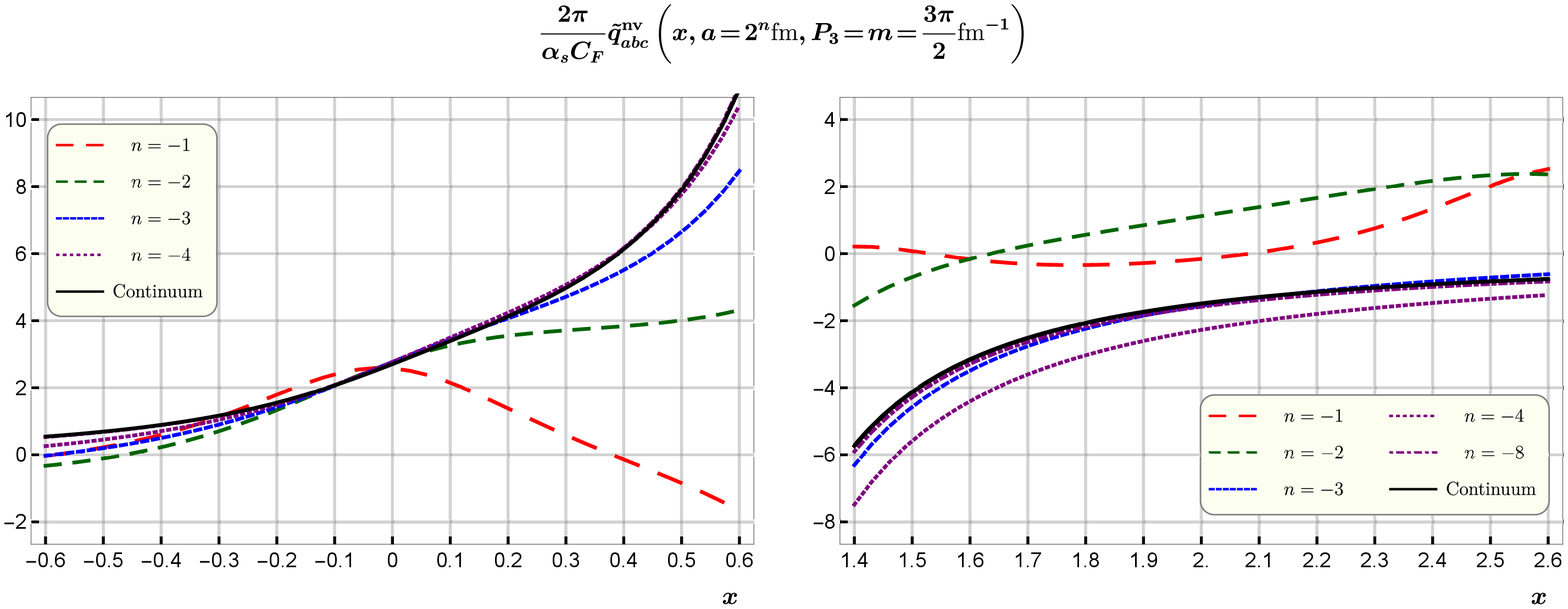}
\par\end{centering}
\begin{centering}
\includegraphics[width=1\textwidth]{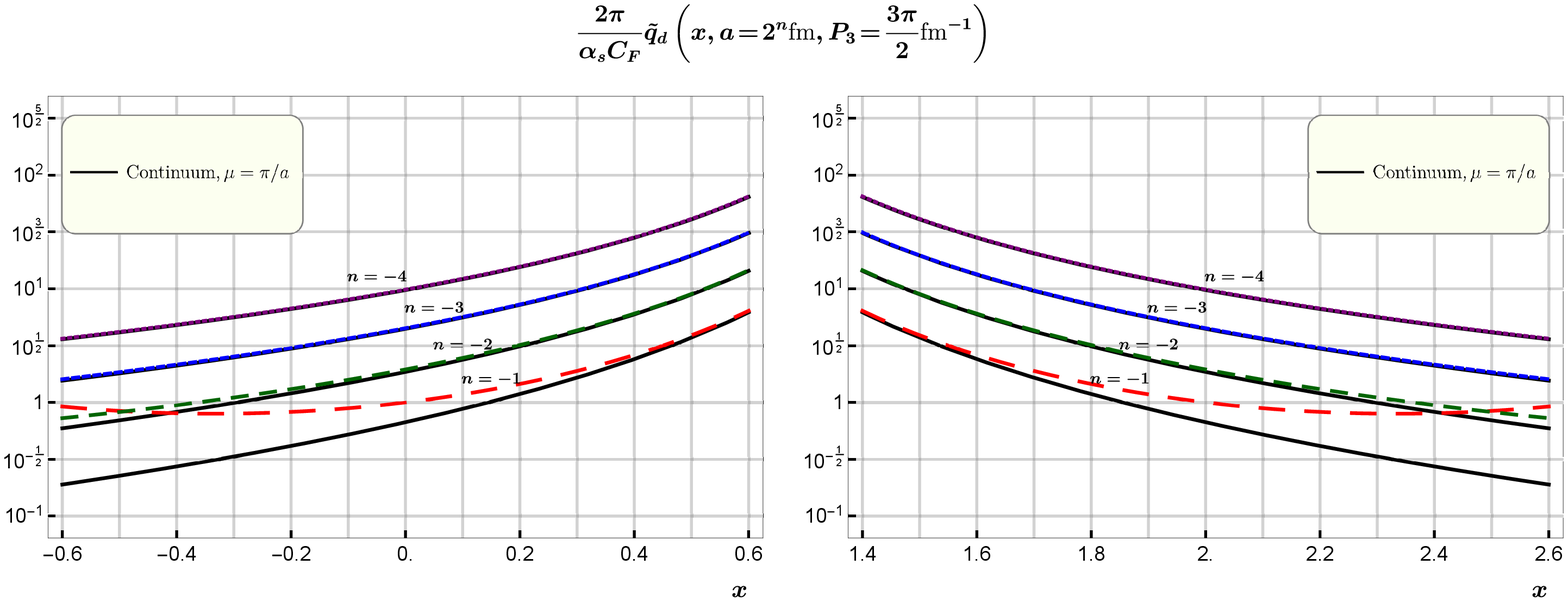}
\par\end{centering}
\centering{}\caption{quasi-PDF in nav\"ie lattice fermion fomulation, with different lattice
spacing. The quasi-PDF in LPT approaches the continuum
quasi-PDF when the lattice spacing is small enough. }
\label{fig:q_abcd}
\end{figure}

The continuum quasi-PDF can be found in Ref.~\cite{Xiong:2013bka}.
The numerical results with different lattice spacing and comparison
to the continuum quasi-PDF are shown in Fig.\ref{fig:q_abcd}. There are
4 different curves for the continuum quasi-PDF $\tilde{q}_{d}\left(x\right)$,
corresponding to the linear UV divergence in continuum quasi-PDF
with the transverse momentum cut-off $\Lambda=\pi/a$. As a consequence,
different lattice spacings result in different continuum quasi-PDFs. The $\tilde{q}_{a,b,c}(x)$
are UV finite, therefore there is only one continuum limit. We use $\tilde{q}^{\text{nv}}$ to denote the quasi-PDF calculated in na\"ive fermion action and $\tilde{q}^{\text{WC}}$ to denote the quasi-PDF calculated in Wilson-Clover fermion action.

\subsection{Numerical results of quasi-PDF in na\"ive lattice fermion action}
From Fig.~\ref{fig:q_abcd} we find that when the lattice spacing is small enough,
the quasi-PDF in na\"ive lattice fermion action ($\tilde{q}^\mathrm{nv}(x)$) closely approaches the continuum
quasi-PDF. For the region $x>1$, the $\tilde{q}_{a,b,c}^{\mathrm{nv}}\left(x\right)$
with $n=-4$ curve appears to be further away from the continuum quasi-PDF
than the $n=-3$ curve. However, this is a coincidence, since as the lattice spacing
further shrinks to $n=-8$, it approaches the continuum quasi-PDF much
closer than $n=-3,-4$. $\tilde{q}_{d}^{\mathrm{nv}}(x)$ is symmetric with
respect to $x=1$ and it converges to the corresponding continuum
quasi-PDF from diagram $d$ faster than $\tilde{q}^{\mathrm{nv}}_{a,b,c}\left(x\right)$. Indeed,
when $a=2^{-3}{\rm fm}$, $q_{d}\left(x\right)$ almost coincides
with the continuum quasi-PDF. The $\mathcal{O}\left(a^{0}\right)$ integrand of diagram $d$ in \eqref{eq:q0d_k_intgrnd} reads
\begin{align}
\mathcal{Q}_{3}^{-2}\mathcal{Q}^{-2} & =\left(\frac{a}{2}\right)^{4}\left[\sin\left(a\frac{P_{3}-k_{3}}{2}\right)\right]^{-2}\sum_{\mu}\left[\sin\left(a\frac{P_{\mu}-k_{\mu}}{2}\right)\right]^{-2},
\end{align}
which is an even function in $\left(1-x\right)$. The $a\rightarrow0$
series expansion gives the continuum integrand $\left(P_{3}-k_{3}\right)^{-2}\left(P-k\right)^{-2}$ with
residual term at $\mathcal{O}\left(a^{2}\right)$, while the difference
between lattice perturbation and continuum integrand of $q^{\mathrm{nv}}_{a,b,c}\left(x\right)$
is of order $\mathcal{O}\left(a^{1}\right)$. Therefore, the contribution
from diagram $d$ approaches its continuum correspondence faster
than the contributions from diagrams $a$, $b$ and $c$.

\begin{figure}
\begin{centering}
\includegraphics[width=1\textwidth]{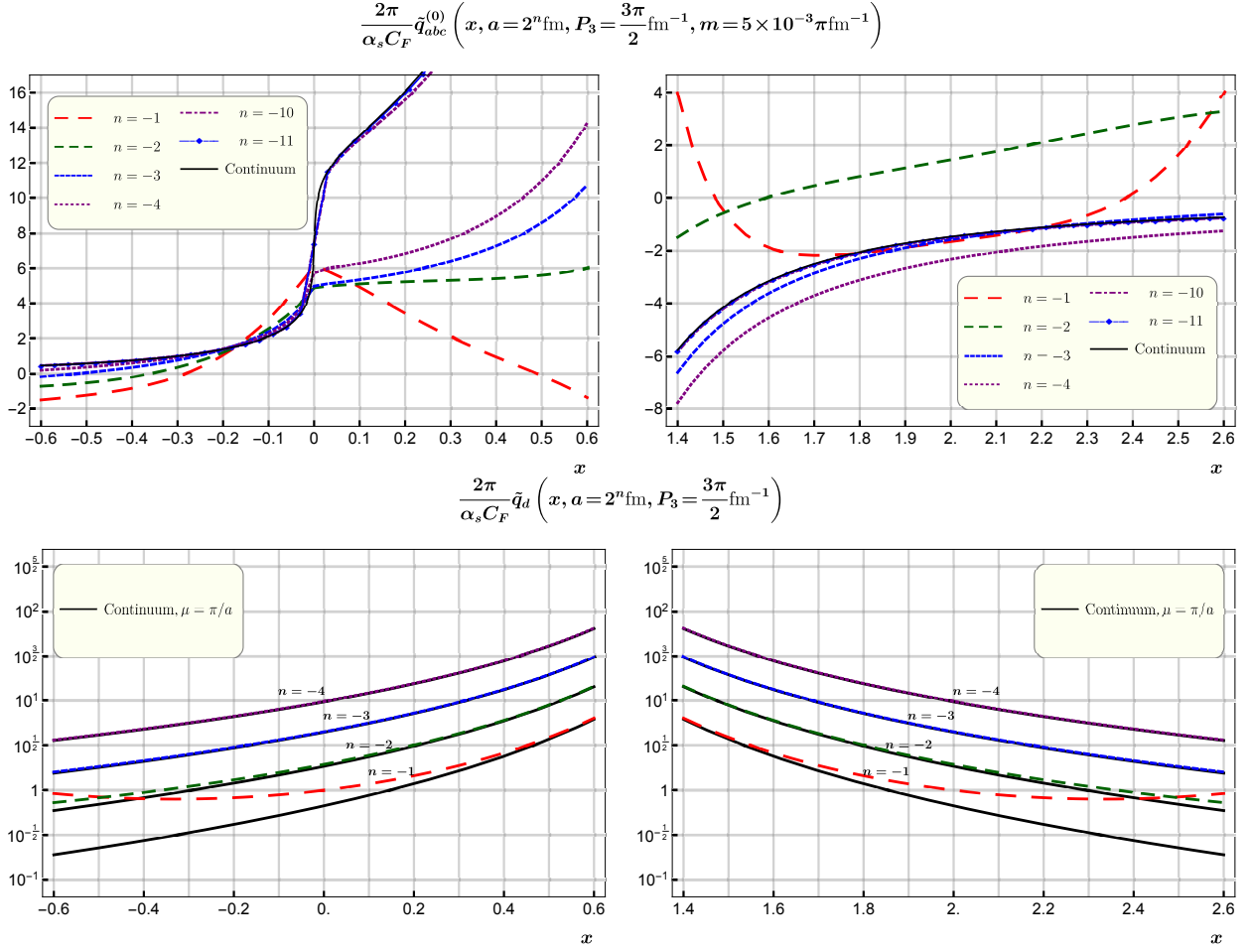}
\par\end{centering}
\begin{centering}
\includegraphics[width=1\textwidth]{Fgd}
\par\end{centering}
\centering{}\caption{Numerical results for the quasi-PDF with  a small quark mass and different lattice
spacings. The quasi-PDF in LPT does not agrees well with
the continuum quasi-PDF when the collinear divergence appears in continuum
quasi-PDF.}
\label{fig:q_abdc_m0}
\end{figure}

\begin{figure}
\begin{centering}
\includegraphics[width=1\textwidth]{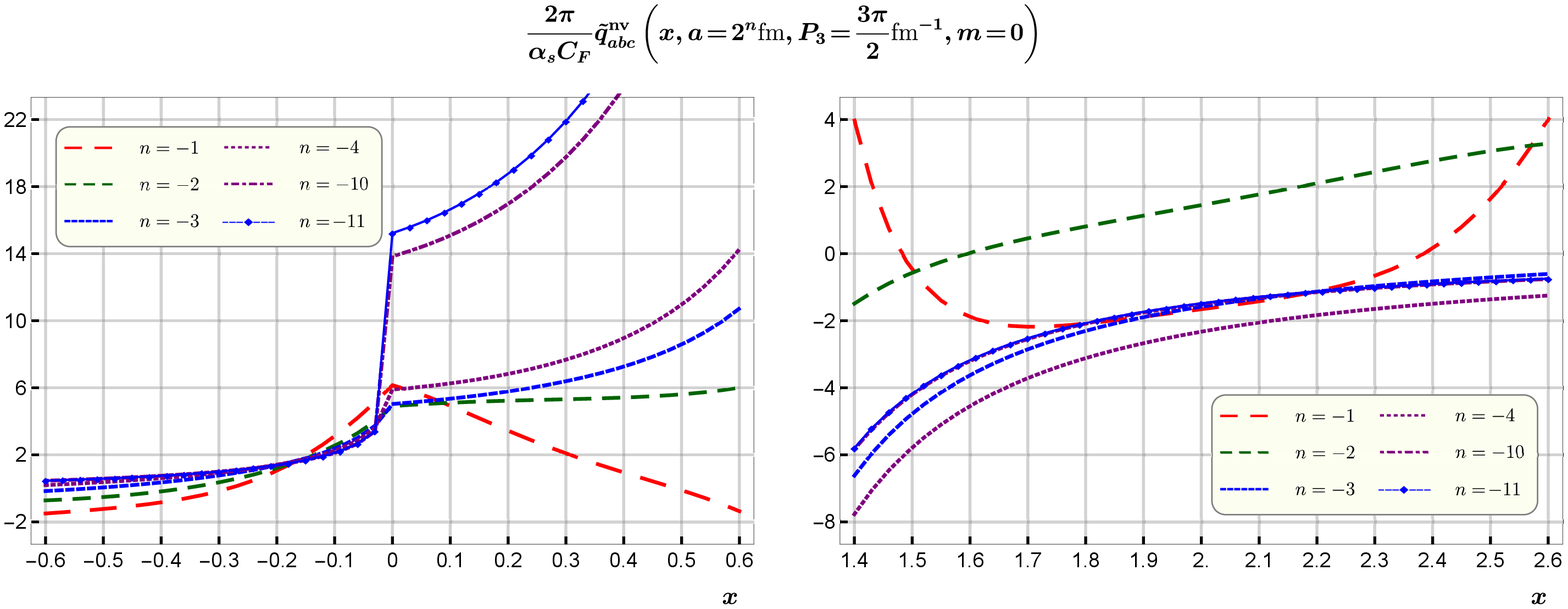}
\par\end{centering}
\caption{quasi-PDF in LPT with massless quarks. There is no collinear
divergence ($\log m^2$) when the lattice spacing is nonzero. 
}
\label{fg:f_0m}
\end{figure}

We also calculated the quasi-PDF with quark mass $m=0.02\pi/L$ to
study the different collinear behavior in LPT-calculated quasi-PDF and
continuum quasi-PDF. The numerical result is shown in Fig.~\ref{fig:q_abdc_m0}.
The lattice perturbation result $\tilde{q}^{\mathrm{nv}}\left(x\right)$ still
coincides with the continuum quasi-PDF in the region $x<0$ (and $x>1$
which is not shown here), because there is no collinear divergence
in the two regions. Consequently, the small quark mass limit and
small lattice spacing limit commute in the lattice perturbation
quasi-PDF. In the region $0<x<1$ there is a significant discrepancy,
because there is a collinear divergence term $\left(1+x^{2}\right)\ln m^{2}/\left(x-1\right)$~\cite{Xiong:2013bka}
while the collinear divergence is absent in LPT-calculated quasi-PDF
due to the finite lattice spacing. However, if we take  $a\rightarrow0$
faster than $m\rightarrow0$ (the purple dot-dashed line, almost overlapping
with the continuum), the collinear divergence actually begins to show up
in the region $0<x<1$. Since there is no collinear divergence both
in the continuum and lattice perturbation calculated quasi-PDF $\tilde{q}_{d}\left(x\right)$,
they always  agree nicely when the lattice spacing $a$ is
smaller than $2^{-3}{\rm fm}$. An extreme case is to set the quark mass
to zero as shown in Fig.~\ref{fg:f_0m}. For nonzero lattice
spacing, the lattice perturbation calculated quasi-PDF (contribution from diagrams
$a$, $b$ and $c$) are free from the collinear divergence in the region $0<x<1$.

\begin{figure}
\centering{}\includegraphics[scale=0.4]{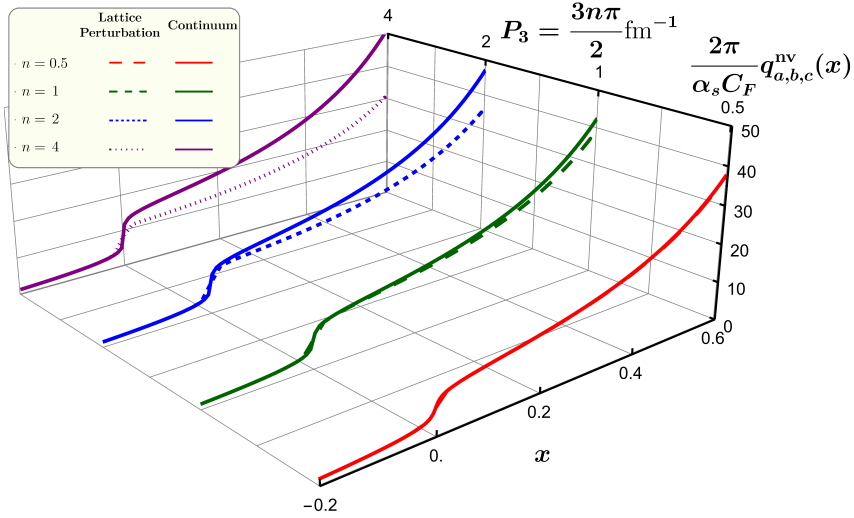}\caption{Numerical results for the quasi-PDF (na\"ive fermion action case) with different quark momentum $P_{3}$
(fixed lattice spacing and quark mass). The increasing $P_{3}$ changes
the order of $aP_{3}^{2}$ and $m$ from $aP_{3}^{2}\ll m$ to $aP_{3}^{2}\gg m$,
resulting in the discrepancy between lattice perturbation and continuum
quasi-PDF.}
\label{fg:nPz_3D}
\end{figure}

It should be noted that the lattice spacing $a$ and quark mass $m$
have different dimensions,  thus it is not meaningful to compare how fast
$a$ and $m$ approach zero directly; rather, one should compare the dimensionless
quantities $aP_{3}$ and $m/P_{3}$. The continuum limit should be understood
as $aP_{3}\ll m/P_{3}$, in other words, $a$ approaches  zero much
faster than $m$. In Fig.\ref{fg:nPz_3D}, we compare the lattice
perturbation quasi-PDF and continuum quasi-PDF (contains collinear divergence $\ln m$) with fixed lattice
spacing $a=2^{-10}~{\rm fm}$ and $m=0.005\pi\text{fm}^{-1}$ but for different quark
momenta $P_{3}=1.5n\pi\,\text{fm}^{-1}$ for $n=0.5,\,1,\,2,\,4$.  The values of
$aP_{3}^{2}/m=0.345,\,1.381,\,5.524,\,22.089$ correspondingly demonstrate the transition
from $aP_{3}\ll m/P_{3}$ to $aP_{3}\gg1$. From the figure we
find that in the region $x<0$ which does not contain any collinear divergence,
the quasi-PDF in LPT always shows good agreement with
the continuum quasi-PDF, while in the region $0<x<1$ where the continuum
quasi-PDF contains a collinear divergence, the discrepancy between lattice
perturbation quasi-PDF and continuum quasi-PDF increases with increasing
$P_{3}$. This is because increasing $P_{3}$ will enhance the effect
from lattice artifacts, i.e. $aP_{3}$ and $aP_{4}$ as analysed analytically in Sec.~\ref{Sec:Cntnm_ClnrDvgnc}. As a consequence,
the implementation of  the target mass correction~\cite{Chen:2016utp} to eliminate
$P_{3}$ power suppressed corrections becomes quite essential, otherwise
one has to pay the price of large lattice artifacts when performing lattice QCDs simulation with large values of $P_{3}$.

\subsection{Numerical results of quasi-PDF in Wilson-Clover lattice fermion action}
\begin{figure}
  \centering{}\includegraphics[width=1\textwidth]{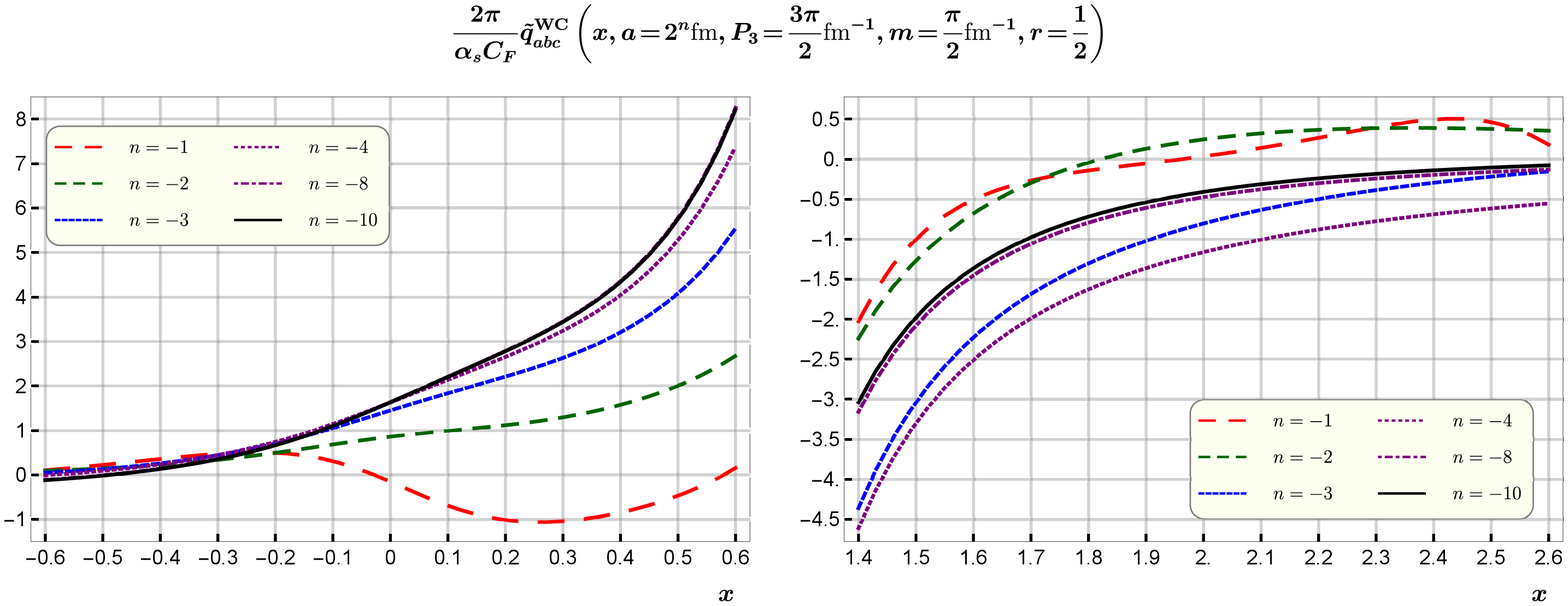}\caption{Numerical results for the  quasi-PDF in Wilson-Clover lattice fermion action with non-zero bare quark mass and different lattice spacing.}
  \label{fg:WCanm1}
  \end{figure}

\begin{figure}
    \centering{}\includegraphics[width=1\textwidth]{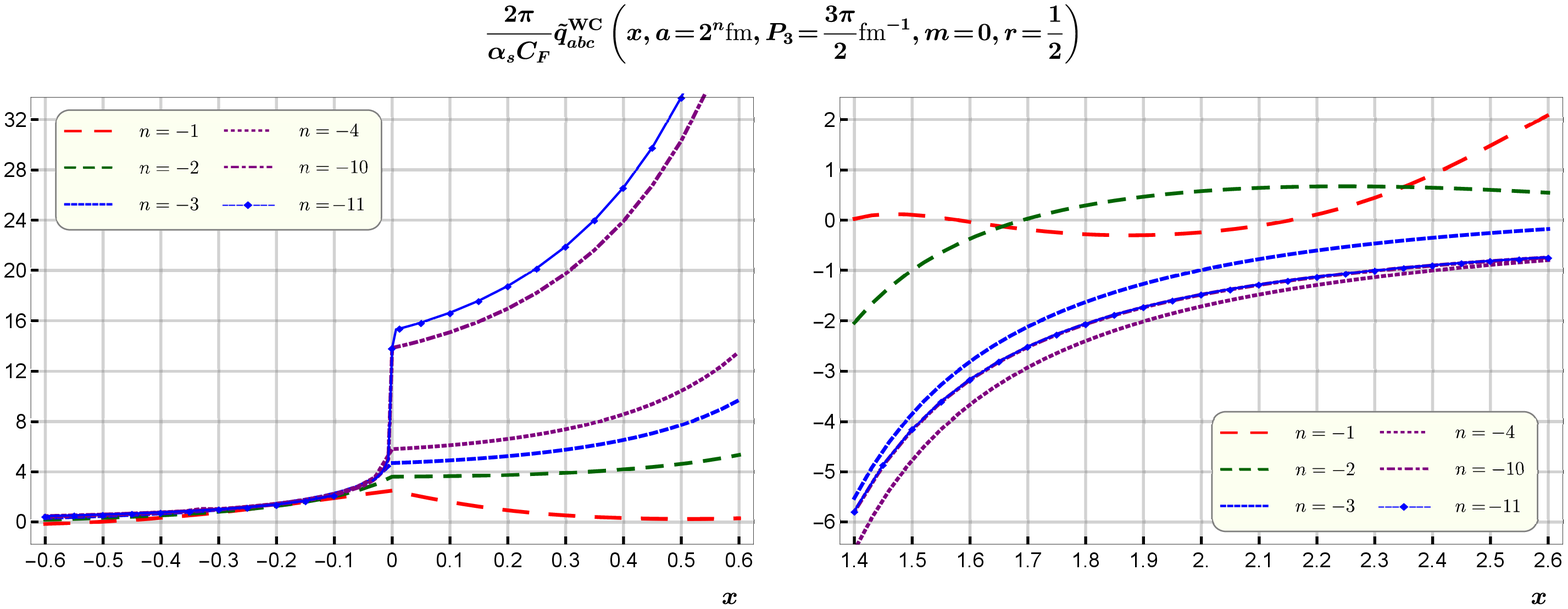}\caption{Numerical results for the  quasi-PDF in Wilson-Clover lattice fermion action with vanishing bare quark mass and different lattice spacing. There is no collinear divergence in the numerical results.}
    \label{fg:WCan0m}
    \end{figure}

We also calculated the quark quasi-PDF in Wilson-Clover action with both massive and massless quark. The Wilson parameter is set as $r=1/2$ and the Clover parameter is chosen to be the leading order in perturbation theory: $c_\mathrm{sw}=1$. The numerical results are shown in Fig. (\ref{fg:WCanm1},\ref{fg:WCan0m}).

\begin{figure}
    \centering{}\includegraphics[width=1\textwidth]{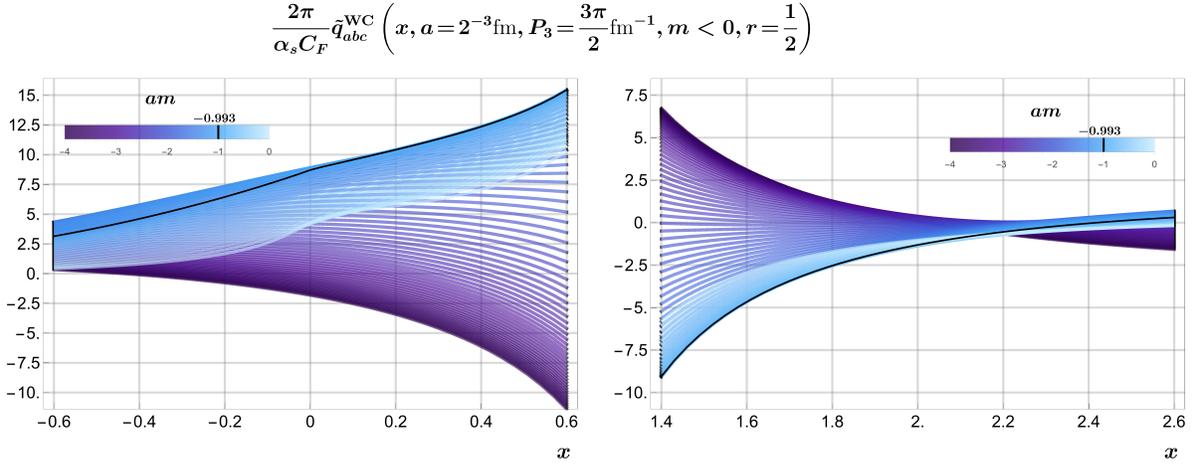}\caption{Numerical results for the  quasi-PDF in Wilson-Clover lattice fermion action with negative bare quark masses $-4\leq am \leq -0.05$, which is around the critical mass $am_0=-0.993$ (black solid line). The collinear divergence is still absent in negative bare quark mass case.}
    \label{fg:WCanNgtvm}
    \end{figure}

{It is worth mentioning that the Wilson-Clover fermion receives an additive mass corrections 
from renormalization, which means the \emph{bare} quark mass (critical mass) can be 
initially negative (which is the case in LQCD calculations) in order to have a vanishing 
renormalized quark mass. We also calculated the quasi-PDF with the bare quark mass set to 
the one-loop critical mass calculated from the quark self-energy diagram (sunset 
diagram and tadpole diagram). We follow Ref.~\cite{Capitani:2002mp}, but 
with $r=1/2$ and $c_\mathrm{SW}=1$)\footnote{{We also compared our bare quark mass $am_0$ 
at $c_\mathrm{SW}=0$,  $r=1$ with the numerical result in Ref.~\cite{Capitani:2002mp}, 
we get  \begin{align*}
 am_0 =& -0.0158473C_F g_s^2-0.309866C_F g_s^2, \;\;\text{Ref.~\cite{Capitani:2002mp}},\\
 am_0 =& -0.0158475C_F g_s^2 -0.309865C_F g_s^2, \;\; \text{this work, setting }c_\mathrm{SW}=0,r=1,
 \end{align*} where the first and second term correspond to the sunset and tadpole diagram, respectively. The relative difference is of the order of $10^{-6}$.}}, the resulting critical mass 
is $am_0\left(\!c_\mathrm{SW}\!=\!1,\!r\!=\!1/2\right) = -0.288C_F g_s^2\left(\mu=\pi/a\right) 
=-0.993$ corresponding to a vanishing one-loop renormalized quark mass. We calculated 
the quasi-PDF with the Wilson-Clover action with bare quark mass ranging from $am=-4$ to 
$ am=-0.05$ so that we can cover the uncertainty from the strong coupling constant. 
The corresponding numerical results are shown in Fig.~\ref{fg:WCanNgtvm}. The collinear 
divergence is still absent in the negative bare quark mass case.}

\begin{figure}
  \begin{centering}
  \includegraphics[width=1\textwidth]{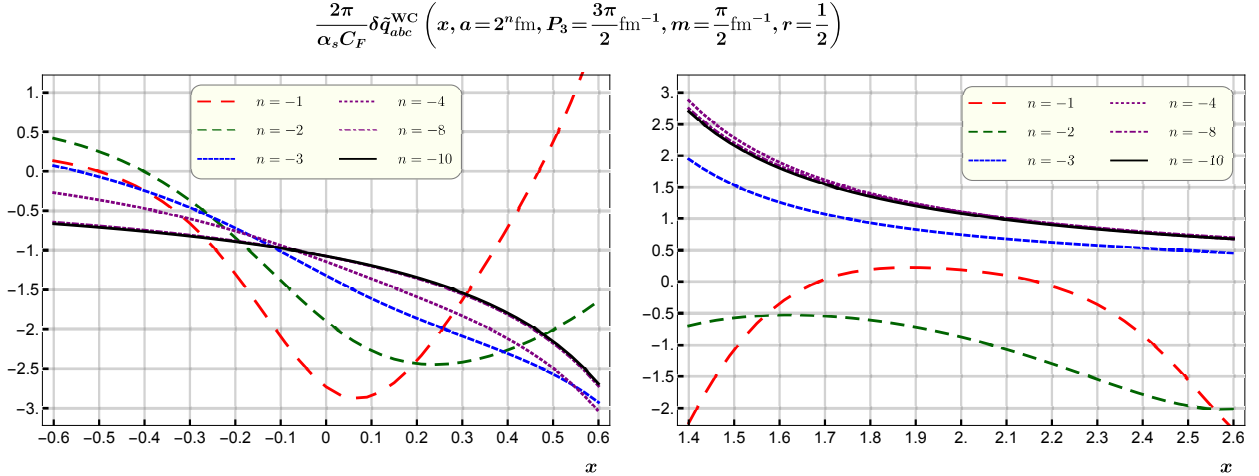}\caption{Numerical results of $\mathcal{O}\left(a^n\right)$ with $n\geq 1$ corrections $\delta\tilde{q}(x,a,P_3)$ to quasi-PDF in na\"ive lattice fermion action. The $\mathcal{O}\left(a^{n}\right)$ with$\;n\geq 1 $
  corrections do not vanish in the  $a\rightarrow0$ limit.}
  \label{fg:fabc_Oa1}
  \par\end{centering}
  \end{figure}

{The difference between the quasi-PDF calculated with the na\"ive fermion and the
Wilson-Clover fermion action is
\begin{align}
\delta \tilde{q}\left(x,P_3\right)=\tilde{q}^{\mathrm{WC}}\left(x,P_3\right)-\tilde{q}^{\mathrm{nv}}\left(x,P_3\right).
\end{align} 
The corresponding numerical results are shown
in Fig.\ref{fg:fabc_Oa1}, we find that $\delta\tilde{q}\left(x,P_3\right)$ converges to a non-vanishing distribution under $a\rightarrow 0$ limit. Since $\lim_{a\rightarrow0} \tilde q^{\mathrm{nv}}(x,P_3) = q(x,P_3)$,  $\tilde q^{\mathrm{WC}}(x,P_3)$ can not recover the continuum quasi-PDF $q\left(x,P_3\right)$. $\delta\tilde{q}(x,P_3)$ comes from the Wilson term 
in the action and a simple series expansion shows that the $k$-integrand of 
$\delta\tilde{q}\left(x,P_3\right)$ is of the order of $\mathcal{O}\left(a^1\right)$ and 
thus can be viewed as $\mathcal{O}\left(a^1\right)$ corrections to the quasi-PDF calculated 
with the na\"ive fermion action ($\tilde q^{\mathrm{nv}}\left(x,P_3\right)$). However, the 
UV region $k_\perp\sim \mathcal{O}\left(a^{-1}\right)$ breaks the expansion and the 
$k_\perp$-integration turns out to be $\mathcal{O}\left(a^0\right)$ and therefore mixes 
with the quasi-PDF calculated with the na\"ive fermion action. This mixing is clearly 
visible in Fig.~\ref{fg:fabc_Oa1} where we see that in the
$a\rightarrow0$ limit, $\delta\tilde{q}\left(x,P_3\right)$ converges
to a non-zero distribution. The mixing behavior is due to the fact that $\delta q(x,P_3)$
contains power divergent UV terms in the $a\rightarrow0$ limit, e.g. after 
$k_4$-integrated out
\begin{align}
\delta\tilde{q}^{\mathrm{WC}}_{bc}(x,P_3)\supset\int_{-\tfrac{\pi}{a}}^{\tfrac{\pi}{a}} \frac{d^2\boldsymbol{k}_\perp}{8\pi^3}\frac{8g_s^2 C_F amr\Pi_-^2\widehat{2k}^2_\perp e^{iaP_4}\widetilde{P-k}_3}{3a\left(\Pi_- - \Pi_+\right)
\displaystyle{\prod_{i=1}^4}\left(\Pi_--z_i\right)\widehat{P-k}_3}.
\end{align}
In order
to extract its UV behavior, we expand its continuum limit at
$\left|\boldsymbol{k}_{\perp}\right|\rightarrow\infty$ , which gives
\begin{equation}
\int \frac{d^2\boldsymbol{k}_\perp}{16\pi^3}\frac{g_s^2C_F mP_3}{P_0^2}\frac{a}{\left|\boldsymbol{k}_\perp\right|}+\cdots
\end{equation}
After integrating out $\boldsymbol{k}_\perp$,the power divergent UV integrand in the 
above equation gives a contribution proportional to $a^{-1}$ which cancels the 
$a^{1}$ prefactor and therefore contributes to the integral at the order $\mathcal{O}\left(a^{0}\right)$.} 
The mixing between higher and lower order in $a$ already appears in the LPT calculation of 
the fermion self-energy with Wilson fermions~\cite{Capitani:2002mp}.The mixing indicates that some non-perturbative matching methods for quasi-PDF is required.

\section{summary}

The PDFs  are one of the most essential non-perturbative quantities in QCD
and they  encode the dynamics of the QCD fundamental degrees of freedom: quarks
and gluons inside a hadron. The PDFs can be measured by high energy scattering
experiments and they are also widely used in  experiments involving
hadrons. The first principal determination of PDFs is still a very
challenging area in QCD research. Large momentum effective field
theory is developed to improve the first principal QCD calculation
of the PDFs and its generalizations.

In this work, we calculated the quark-in-quark quasi-PDF $\tilde{q}\left(x,P_3\right)$ with na\"ive fermion action ($\tilde{q}^{\mathrm{nv}}(x)$) and quasi-PDF with Wilson-Clover fermion action ($\tilde{q}^{\mathrm{WC}}(x)$).
We showed  analytically that the
$\boldsymbol{k}_{\perp}$-unintegrated $\tilde{q}^{\mathrm{nv}}(x)$ exactly reduces to the continuum quasi-PDF in the zero lattice
spacing limit. From the analytical $\boldsymbol{k}_{\perp}$-unintegrated $\tilde{q}^{\mathrm{nv}}(x)$, we found that the collinear divergence
is absent in the LPT calculated quasi-PDF at finite lattice spacing.
We also compared the collinear behavior of $\tilde{q}^{\mathrm{nv}}(x)$
and the continuum $\boldsymbol{k}_{\perp}$-integrated quasi-PDF numerically, and
found that the limit of massless quark and zero lattice spacing
do not commute. Our findings demonstrate that the \emph{proper} limit to recover the collinear
divergence of the continuum quasi-PDF should be $aP_{3}^{2}\approx m$ and $m \ll P_{3}$, while the limit to fully recover
the continuum quasi-PDF is $aP_{3}^{2}\ll m\ll P_{3}$. These two conditions are based on perturbative calculations, therefore they should not be applied to non-perturbative lattice calculations which do not contain the collinear divergence because of the non-perturbative effects. The difference between $\tilde{q}^{\mathrm{WC}}(x)$ and $\tilde{q}^{\mathrm{nv}}(x)$ ($\delta\tilde q\left(x,P_3\right)=\tilde{q}^{\mathrm{WC}}(x)-\tilde{q}^{\mathrm{nv}}(x)$ can be viewed as a $\mathcal{O}\left(a^{1}\right)$ correction to $\tilde{q}^{\mathrm{nv}}(x)$) is due to the Wilson term and it contains UV power divergent integrands. These UV power divergent integrands render $\mathcal{O}\left(a^1\right)$ corrections that mix with $\tilde{q}^\mathrm{nv}(x)$.
This mixing indicates that a non-perturbative matching is required, and will be a subject of future research.

\vspace{0.3 cm}
{\noindent \it Acknowledgments}

We thank Evan Berkowitz for useful discussions. X.X. also thanks Jian-Hui Zhang and Yong Zhao for discussions. We acknowledge financial support from the Deutsche Forschungsgemeinschaft (Sino-German CRC 110). The work of UGM was also supported by the Chinese Academy of Sciences (CAS) President's International Fellowship Initiative (PIFI) (Grant No.2017VMA0025).

\allowdisplaybreaks[1]

\appendix
\section{Expressions For $k_4$-Integration of quasi-PDF in na\"ive Lattice Fermion}\label{Oa0_k4_intgrtn_exprsn}
The $k_4$-integration of quasi-PDF calculated with na\"ive fermion action is given by the follows and in which $k_3=xP_3$
\begin{align}
& \tilde{q}_{a}^{\mathrm{nv}}\left(x\right)\nonumber\\
= & P_{3}g_{s}^{2}C_{F}\int_{-\frac{\pi}{a}}^{\frac{\pi}{a}}\frac{d^{2}\boldsymbol{k}_{\perp}}{\left(2\pi\right)^{2}}\;\left[\frac{a\sqrt{\Gamma_{-}}e^{iaP_{4}}\left(a^{2}\left(-\widetilde{k+P}_{3}^{2}\!+\!\widetilde{\boldsymbol{k}}_{\perp}^{2}\right)\!+\!2\right)\!+\!a^{2}\Gamma_{-}e^{2iaP_{4}}\!+\!1}{16\pi^{3}a\left(\Gamma_{-}-\Gamma_{+}\right)\left(a\Pi_{-}-\sqrt{\Gamma_{-}}\right)\left(a\Pi_{+}-\sqrt{\Gamma_{-}}\right)}\right.\nonumber\\
 & \left.\boldsymbol{+}\left\langle \sqrt{\Gamma_{-}}\rightarrow\!-\!\sqrt{\Gamma_{-}}\right\rangle \vphantom{\frac{\widetilde{P}_{3}^{2}}{\sqrt{\Gamma_{-}}}}\right]+\frac{\Pi_{-}\left(a^{4}\Pi_{-}e^{iaP_{4}}\left(-\widetilde{k+P}_{3}^{2}+\widetilde{\boldsymbol{k}}_{\perp}^{2}\right)+\left(1+a^{2}\Pi_{-}e^{iaP_{4}}\right){}^{2}\right)}{8\pi^{3}a\left(\Pi_{-}-\Pi_{+}\right)\left(\Gamma_{-}-a^{2}\Pi_{-}^{2}\right)\left(\Gamma_{+}-a^{2}\Pi_{-}^{2}\right)}\nonumber\\
 & +\mathcal{F}_{1}\left(\mathcal{X},\mathcal{Y}\right)+\frac{4m^{2}}{\widehat{2k}_{3}\widehat{2P}_{3}}\mathcal{F}_{1}\left(-a^{2}\widetilde{k+P}_{3}^{2},a^{2}\widehat{\boldsymbol{k}}_{\perp}^{2}-10\right)+\frac{i\widehat{2P}_{4}\left(a^{4}\Pi_{-}^{2}-1\right)}{a^{3}\Pi_{-}\widehat{2k}_{3}\widehat{2P}_{3}}\nonumber\\
 & \times\mathcal{F}_{1}\left(a^{2}\widetilde{k\!+\!P}_{3}^{2},-a^{2}\widehat{\boldsymbol{k}}_{\perp}^{2}\!+\!6\right)\!+\!\mathcal{F}_{2}\left(\sqrt{\Gamma_{-}},a^{2}\widetilde{k\!+\!P}_{3}^{2}\right)+\frac{4m^{2}}{\widehat{2k}_{3}\widehat{2P}_{3}}\mathcal{F}_{2}\left(\sqrt{\Gamma_{-}},-a^{2}\widetilde{k\!+\!P}_{3}^{2}\right)\nonumber\\
 & +\left\langle \!\mathcal{F}_{2}\left(\!-\sqrt{\Gamma_{-}},a^{2}\widetilde{k\!+\!P}_{3}^{2}\!\right)\!+\!\frac{4m^{2}}{\widehat{2k}_{3}\widehat{2P}_{3}}\mathcal{F}_{2}\left(\!-\sqrt{\Gamma_{-}},-a^{2}\widetilde{k\!+\!P}_{3}^{2}\!\right)\!\right\rangle \!-\!\frac{a^{2}\widetilde{k\!+\!P}_{3}^{2}\!+\!a^{2}\widetilde{\boldsymbol{k}}_{\perp}^{2}\!-\!2}{16\pi^{3}a^{8}\Gamma_{-}\left(\!\Gamma_{-}\!-\!\Gamma_{+}\!\right){}^{3}\widehat{2P}_{3}}\nonumber\\
 & \times\left[\frac{i\widehat{2k}_{3}\widehat{2P}_{4}e^{iaP_{4}}\left(-2a^{4}\Gamma_{-}^{2}\!+\!a^{3}\sqrt{\Gamma_{-}}\left(\Pi_{-}\!+\!\Pi_{+}\right)\left(a^{2}\Gamma_{-}\!-\!1\right)\!+\!2e^{2iaP_{4}}\right)}{\left(\!\sqrt{\Gamma_{-}}\!-\!a\Pi_{-}\!\right){}^{2}\left(\!\sqrt{\Gamma_{-}}\!-\!a\Pi_{+}\!\right){}^{2}\left(\!a^{2}\Gamma_{-}\!-\!1\!\right){}^{-2}}\boldsymbol{+}\left\langle \sqrt{\Gamma_{-}}\rightarrow-\sqrt{\Gamma_{-}}\right\rangle \right]\nonumber\\
 & +\left\{ \frac{a^{-7}\Gamma_{-}^{-1}\left(\Gamma_{-}\!-\!\Gamma_{+}\right){}^{-3}\widehat{2k}_{3}\widehat{2P}_{4}}{16\pi^{3}\widehat{2P}_{3}\left(\sqrt{\Gamma_{-}}\!-\!a\Pi_{-}\right){}^{2}\left(\sqrt{\Gamma_{-}}\!-\!a\Pi_{+}\right){}^{2}}\left[-i\sqrt{\Gamma_{-}}\left(a^{2}\Gamma_{-}\!-\!1\right){}^{2}\right.\right.\nonumber\\
 & \times\left(2a^{3}\sqrt{\Gamma_{-}}\left(\Pi_{-}\!+\!\Pi_{+}\right)\left(e^{2iaP_{4}}-1\right)\!+\!3a^{2}\Gamma_{-}\!-\!\left(a^{2}\Gamma_{-}\!+\!3\right)e^{4iaP_{4}}+1\right)\nonumber\\
 & \left.-ia^{4}\Gamma_{-}^{5/2}\left(a^{2}\Gamma_{-}\!-\!2\right)\left(a^{2}\Gamma_{-}\!-\!1\right)e^{2iaP_{4}}\!-\!i\sqrt{\Gamma_{-}}\left(2a^{4}\Gamma_{-}^{2}\!-\!a^{2}\left(4\Gamma_{-}\!+\!\Gamma_{+}\right)\!+\!3\right)e^{2iaP_{4}}\right]\nonumber\\
 & \left.\!+\!\left\langle \sqrt{\Gamma_{-}}\rightarrow\!-\!\sqrt{\Gamma_{-}}\right\rangle \vphantom{\frac{\widehat{P}_{4}}{\left(\sqrt{\Gamma_{-}}\right){}^{2}}}\right\} +\left\{ \frac{a^{-7}\Gamma_{-}^{-1}\left(\Gamma_{-}\!-\!\Gamma_{+}\right){}^{-3}\widehat{2k}_{3}\left(\widehat{2k}_{3}\widehat{2P}_{3}+4m^{2}\right)}{16\pi^{3}\widehat{2P}_{3}\left(\sqrt{\Gamma_{-}}\!-\!a\Pi_{-}\right){}^{2}\left(\sqrt{\Gamma_{-}}-a\Pi_{+}\right){}^{2}}\right.\nonumber\\
 & \times\left[\vphantom{\left(\left(a^{4}\Gamma_{-}^{2}\right){}^{2}\right)}-4a^{6}\Gamma_{-}^{3}-4a^{2}\Gamma_{-}e^{4iaP_{4}}+e^{2iaP_{4}}\left(a^{5}\Gamma_{-}^{3/2}\left(\Pi_{-}\!+\!\Pi_{+}\right)\left(a^{4}\Gamma_{-}^{2}\!+\!3\right)\!-\!2\left(a^{4}\Gamma_{-}^{2}\!+\!1\right){}^{2}\right)\right.\nonumber\\
 & \left.\left.\!+\!a^{3}\sqrt{\Gamma_{-}}\left(\Pi_{-}\!+\!\Pi_{+}\right)\left(3a^{4}\Gamma_{-}^{2}\!+\!1\right)\right]\!+\!\left\langle \sqrt{\Gamma_{-}}\rightarrow\!-\!\sqrt{\Gamma_{-}}\right\rangle \vphantom{\frac{\widehat{P}_{4}}{\left(\sqrt{\Gamma_{-}}\right){}^{2}}}\right\} \nonumber\\
 & \boldsymbol{+}\left\langle \frac{i\Gamma_{-}\widehat{2P}_{4}\left(a^{2}\Gamma_{-}\left(a^{2}\Gamma_{-}\!-\!3\right)\!+\!2\right)\!-\!a\left(\Pi_{-}\!+\!\Pi_{+}\right)
 \left(a^{4}\Gamma_{-}^{2}\!+\!3\right)\left(\widehat{2k}_{3}\widehat{2P}_{3}\!+\!4m^{2}\right)}{8\pi^{3}a^{3}
 \left(\Gamma_{-}\!-\!\Gamma_{+}\right){}^{3}\widehat{2P}_{3}\left(a\Pi_{-}\!+\!\sqrt{\Gamma_{-}}\right){}^{2}
 \left(a\Pi_{+}\!+\!\sqrt{\Gamma_{-}}\right){}^{2}\Gamma_{-}^{-\frac{1}{2}}\widehat{2k}_{3}^{-1}e^{-2iaP_{4}}}\right\rangle ,\label{eq:fa0k4Inted}
\end{align}
where $\left.\langle \sqrt{\Gamma_-}\rightarrow-\sqrt{\Gamma_-} \right\rangle$ means performing the replacement to the previous term within the same $\left\{\cdots\right\}$ or $\left[\cdots \right]$. The functions $\mathcal{F}_{1,2}$ are defined as
\bseq
\begin{align}
\mathcal{F}_{1}\left(\mathcal{X},\mathcal{Y}\right)=&\frac{a\Pi_{-}^{3}\widehat{2k}_{3}^{2}
\left(-a^{2}\Pi_{-}e^{iaP_{4}}\left(\mathcal{X}+\mathcal{Y}\right)+a^{4}\Pi_{-}^{2}e^{2iaP_{4}}+1\right)}
{4\pi^{3}\left(\Pi_{-}-\Pi_{+}\right)\left(\Gamma_{-}-a^{2}\Pi_{-}^{2}\right){}^{2}\left(\Gamma_{+}-a^{2}\Pi_{-}^{2}\right){}^{2}}
\end{align}
\begin{align}
\mathcal{F}_{2}\left(\mathcal{X},\mathcal{Y}\right)=&\frac{-\widehat{2k}_{3}^{2}e^{iaP_{4}}\left(3a^{6}\Gamma_{-}^{3}\!+\!
\left(a^{4}\Gamma_{-}^{2}\!+\!3\right)e^{2iaP_{4}}\!+\!a^{2}\Gamma_{-}-2a^{3}\mathcal{X}\left(\Pi_{-}\!+\!\Pi_{+}\right)
\left(a^{4}\Gamma_{-}^{2}\!+\!1\right)\right)}{16\pi^{3}a^{6}\mathcal{X}\left(\Gamma_{-}\!-\!\Gamma_{+}\right){}^{3}
\left(\mathcal{X}\!-\!a\Pi_{-}\right){}^{2}\left(\mathcal{X}\!-\!a\Pi_{+}\right){}^{2}
\left(a^{2}\widetilde{\boldsymbol{k}}_{\perp}^{2}\!+\!2\!-\!\mathcal{Y}\right)^{-1}}
\end{align}
\eseq
The terms inside ${\left\langle \;\right\rangle }$
are the contributions from the unphysical quark pole $z_{q}^{-}$
and they are of order $\mathcal{O}\left(a^{2}\right)$.

It is straight forward to verify that diagram $b$ and $c$ give an identical
contribution, which reads
\begin{align}
\tilde{q}_{b/c}^{\mathrm{nv}}\left(x\right)\!=\! & \int_{-\frac{\pi}{a}}^{\frac{\pi}{a}}\!d^{2}\boldsymbol{k}_{\perp}\,
\frac{g_{s}^{2}C_{F}}{16\pi^{3}}\frac{P_{3}e^{iaP_{4}}\widetilde{k\!+\!P}_{3}}{a\widehat{2P}_{3}\widehat{k\!-\!P}_{3}}
\left\{\frac{2\Pi_{-}\!\left(\!a^{3}\Pi_{-}\!\left(\widehat{2k}_{3}\widehat{2P}_{3}\!-\!4m^{2}\!\right)\!
+\!i\widehat{2P}_{4}\left(a^{4}\Pi_{-}^{2}\!-\!1\right)\right)}{-\left(\Pi_{-}\!-\!\Pi_{+}\right)\!
\left(a^{2}\Pi_{-}^{2}\!-\!\Gamma_{-}\right)\!\left(a^{2}\Pi_{-}^{2}\!-\!\Gamma_{+}\right)}\right.\nonumber \\
& \left.+\left[\frac{a^{2}\sqrt{\Gamma_{-}}\left(\widehat{2k}_{3}\widehat{2P}_{3}\!-\!4m^{2}\right)
+i\widehat{2P}_{4}\left(1\!-\!a^{2}\Gamma_{-}\right)}{\left(\Gamma_{-}\!-\!\Gamma_{+}\right)\left(a\Pi_{-}\!+\!\sqrt{\Gamma_{-}}\right)\left(a\Pi_{+}\!+\!\sqrt{\Gamma_{-}}\right)}+{\left\langle \sqrt{\Gamma_{-}}\rightarrow-\sqrt{\Gamma_{-}}\right\rangle }\right]\right\}.\label{eq:fbc0k4Inted}
\end{align}

The $k_{4}$ Integration of diagram $d$ gives
\begin{align}
\tilde{q}_{d}\left(x\right)= & \frac{g_{s}^{2}C_{F}}{8\pi^{3}}\int_{-\frac{\pi}{a}}^{\frac{\pi}{a}}d^{2}\boldsymbol{k}_{\perp}\,\frac{P_{3}e^{iaP_{4}}}{a\left(\Pi_{-}-\Pi_{+}\right)\widehat{P-k}_{3}^{2}}.\label{eq:fd0k4Inted}
\end{align}

We also have checked that the $a\rightarrow 0$ limit of above the $\boldsymbol{k}_\perp$-integrands coincide with the $\boldsymbol{k}_\perp$
unintegrated quasi-PDF calculated directly in the continuum.

\end{document}